\documentclass[12pt]{elsarticle}
\usepackage[dvipsnames]{xcolor}
\RequirePackage[T1]{fontenc}
\RequirePackage{mathptmx}
\usepackage{fullpage}
\usepackage{slashed}
\usepackage{graphics,bm}
\usepackage{epsfig}
\usepackage{graphicx}
\usepackage{amsmath}
\usepackage{amssymb}
\usepackage{bbold}
\usepackage{wrapfig}
\usepackage{color}
\usepackage{hyperref}
\usepackage{mathtools}
\usepackage{cancel}
\usepackage{multirow}
\usepackage{upgreek}
\usepackage{subcaption}
    \usepackage[vertfit]{breakurl} 

\journal{}
\DeclareMathOperator{\Tr}{Tr}
\begin{document}
\begin{frontmatter}
\title{Topological susceptibility and fourth cumulant in a uniform magnetic field }
\author{Prabal Adhikari}
\ead{adhika1@stolaf.edu}
\address{Physics Department, Faculty of Natural Sciences and Mathematics, St. Olaf College, 1520 St. Olaf Avenue, Northfield, MN 55057, United States}
\date{\today}
\begin{abstract}
We study the topological susceptibility and fourth cumulant of the QCD vacuum in a background magnetic field using three-flavor chiral perturbation theory ($\chi$PT) for arbitrary quark masses and $n$-flavor $\chi$PT with degenerate quark masses. We find that the enhancement of the topological susceptibility is larger in the three-flavor $\chi$PT compared to two-flavor $\chi$PT. Additionally, in comparing the fourth cumulant, we find that its suppression is comparable for magnetic fields, $eH\lesssim 0.8m_{\pi}^{2}$, and weaker for larger magnetic fields in three-flavor $\chi$PT with its enhancement beginning at a significantly lower critical magnetic field compared to two-flavor $\chi$PT. We also find that the enhancement of the topological susceptibility in $n$-flavor $\chi$PT with degenerate quarks is significantly larger and the suppression of the topological cumulant significantly greater at weak fields with the critical magnetic field pushed out to larger magnetic fields compared to both two and three-flavor $\chi$PT.
\end{abstract}
\end{frontmatter}
\section{Introduction}
The QCD vacuum possesses topological properties due to the axial anomaly~\cite{Crewther:1977ce,Witten:1979vv,Veneziano:1979ec,DiVecchia:1980yfw} that arises via instantons~\cite{tHooft:1976rip, tHooft:1976snw}. Their properties in the QCD vacuum are characterized by topological cumulants such as the topological susceptibility and the fourth cumulant. The former was first studied in the context of the 't Hooft large-$N_{c}$ limit of QCD, where $g^{2}N_{c}$ is kept at order one while $N_{c}\rightarrow\infty$, which implies that the strong coupling constant $g\sim\mathcal{O}(1/\sqrt{N_{c}})$ -- in three-flavor $\chi$PT at large $N_{c}$, the tree level mass of the $\eta'$ meson is given by the Witten-Veneziano formula~\cite{Witten:1979vv,Veneziano:1979ec}
\begin{equation}
\begin{split}
\mathring{m}_{\eta'}^{2}=\frac{2n\chi_{t}^{\infty}}{f^{2}}+\mathring{m}_{\pi}^{2}\ ,
\end{split}
\end{equation}
where $n$ is the number of flavors, $\chi_{t}^{\infty}$ is the topological susceptibility and $f$ is the pion decay constant. In the absence of the axial anomaly at large $N_{c}$, the topological susceptibility, which is $\mathcal{O}(N_{c}^{0})$ vanishes giving rise to a ninth pseudo-Goldstone boson, the $\eta'$, the mass of which is degenerate with the mass of the pions. Alternatively, with $U(1)_{A}$ broken the masses are non-degenerate with corrections of $\mathcal{O}(N_{c}^{-1})$. In real QCD, there are no parity partners associated with the pions suggesting that $U(1)_{A}$ is broken. However, spontaneous breaking of $U(1)_{A}$ is ruled out by the absence of a corresponding isosinglet pseudoscalar expected to possess a mass less than the Weinberg bound of $\sqrt{3}m_{\pi}$~\cite{Weinberg:1975ui}, which is absent in the QCD spectrum. The masses of the closest candidates, the $\eta$ and the $\eta'$, are significant larger suggesting that the axial anomaly is strongly broken by the QCD vacuum. 

The effects of the axial anomaly in QCD can be studied by introducing a $\theta$-term in the QCD Lagrangian,
\begin{equation}
\begin{split}
\mathcal{L}_{\rm QCD}=&-\frac{1}{4}G_{\mu\nu}^{a}G_{a\mu\nu}-\frac{g^{2}\theta }{32\pi^{2}}\tilde{G}^{a\mu\nu}G_{\mu\nu}^{a}+\bar{q}\left (i\slashed{D}-M \right )q\ ,
\end{split}
\end{equation}
where $G^{a}_{\mu\nu}$ is the gluon field strength tensor, $\tilde{G}^{a\mu\nu}$ is the dual field strength tensor, $q$ is the quark field, $M$ is the quark mass matrix and the covariant derivative $\slashed{D}_{\mu}=\slashed{\partial}_{\mu}-ig\slashed{A}^{a}_{\mu}\tfrac{\lambda^{a}}{2}-ieQ\slashed{A}^{\rm ext}$, where we have introduced a $U(1)_{\rm em}$ vector potential since the objective of this paper is to study the effects of a uniform, background magnetic field on the topological susceptibility and the fourth cumulant. While the QCD Lagrangian is symmetric under $U(1)_{A}$ rotations, i.e. $q\rightarrow e^{-i\alpha\gamma_{5}}q$, in the chiral limit, the path integral measure in the QCD partition function, $Z$,
\begin{equation}
\begin{split}
Z=\int \mathcal{D}A\mathcal{D}q\mathcal{D}\bar{q}\exp\left [i\int d^{4}x\mathcal{L}_{\rm QCD} \right ]\ ,
\end{split}
\end{equation}
 is not since the quark and anti-quark field measures transform as~\cite{Fujikawa:1979ay} 
\begin{align}
\mathcal{D}\bar{q}\mathcal{D}q\rightarrow\exp\left [-i\int d^{4}x\frac{g^{2}\alpha n}{16\pi^{2}}\tilde{G}^{a\mu\nu}G_{a\mu\nu} \right ]\mathcal{D}\bar{q}\mathcal{D}q\ .
\end{align}
Choosing $\alpha$ to be $-\frac{\theta}{2n}$ removes the $\theta$-dependence from the dual field strength tensor while transforming the diagonal quark mass matrix from $M\rightarrow e^{-i\theta/n}M$~\cite{srednicki2007quantum}. 

We can study the topological susceptibility and fourth cumulant for small quark masses and weak magnetic fields in $n$-flavor QCD using $\chi$PT, which is the effective field theory that characterizes the interactions of the $n^{2}-1$ pseudo-Goldstone modes. The results are model-independent as long as the magnetic fields ($\sqrt{eH}$) and masses are small compared to $4\pi f_{\pi}$, which is the characteristic scale comparable to the mass of the lightest hadrons, that arises in next-to-leading order calculations in $\chi$PT. Recently, we studied the analogous problem in two-flavor $\chi$PT~\cite{Adhikari:2021xra} while also studying the $\theta$-vacuum topological cumulants since the full free energy is calculable up to $\mathcal{O}(p^{4})$. Unfortunately, this is not possible in three-flavor $\chi$PT since there is no closed-form solution for the ground state, which can be only be calculated recursively~\cite{Lu_2020}. As such we will calculate the QCD vacuum topological susceptibility and fourth cumulant of the $\theta=0$ vacuum in a background magnetic field in the three-flavor case and compare our results with that from the two-flavor case~\cite{Adhikari:2021xra}. Furthermore, we also study the effect of a background magnetic field on the topological cumulants of $n$-flavor $\chi$PT assuming degenerate quark masses.

The primary motivation for focusing on the effect of magnetic fields on topological cumulants is their relevance to a wide array physical systems including magnetars and quark-gluon plasma, which are generated in the early cosmological period of the universe and in heavy ion collisions, where the recent focus has primarily been on the chiral magnetic effect. It is anticipated that in the simultaneous presence of both an electric and a magnetic field, an anomalous current flows in the direction of the external magnetic field with the chiral imbalance generate by an axial chemical potential or an electric field. While there has been significant attention paid to the chiral magnetic effect in heavy ion collisions, until a recent letter~\cite{Adhikari:2021lbl} there had not been much attention paid to the effect on the topological cumulants by the external magnetic field or an external magnetic in the presence of a $\theta$ angle~\cite{Adhikari:2021xra}. We found sum rules relating the shift of the topological susceptibility to that of the quark condensates and the fourth cumulant to that of the quark condensates and susceptibilities. In this paper, we will focus on the topological susceptibility and fourth cumulant in three-flavor $\chi$PT with non-degenerate quark masses and $n$-flavor $\chi$PT with degenerate quarks in the presence of a uniform magnetic background. Furthermore, we will compare the first two cumulants of the $\theta=0$ vacuum in two and three-flavor $\chi$PT. The results presented here will provide future lattice practitioners with relevant model-independent results for such a comparison. For a review of lattice calculations at zero magnetic field, see Ref.~\cite{VICARI200993}. 

The paper is organized as follows: we begin with the calculation of the three-flavor free energy in Section~\ref{sec:3f} and use it to calculate the topological susceptibility and fourth cumulant in Section~\ref{sec:top}, and the chiral condensates and chiral susceptibilities in Section~\ref{sec:cc}. In Section~\ref{sec:sumrules}, we review the sum rules in three-flavor $\chi$PT~\cite{Adhikari:2021lbl} that relate the shift in the topological susceptibility to that of the quark condensate and the shift of the fourth cumulant to the quark condensates and susceptibilities and in Section~\ref{sec:nf}, we calculate the topological cumulants in $n$-flavor $\chi$PT with degenerate quarks. In Section~\ref{sec:discussion}, we plot our analytical results for the topological cumulants and compare them to the two-flavor results from Ref.~\cite{Adhikari:2021xra} and conclude in Section \ref{sec:conclusion}. We list some useful finite magnetic field integrals in \ref{sec:integrals} and provide one-loop renormalized expression for the pion decay constants, pion masses and kaon masses in two and three-flavor $\chi$PT in \ref{sec:renormalized}. 
\section{Three-flavor $\chi$PT}
\label{sec:3f}
\noindent
We begin with the leading order three-flavor $\chi$PT  Lagrangian~\cite{Gasser:1984gg,scherer2011primer}
\begin{equation}
\begin{split}
\mathcal{L}_{2}&=-\frac{1}{4}F_{\mu\nu}F^{\mu\nu}+\frac{f^{2}}{4}\Tr\left [\nabla_{\mu}\Sigma(\nabla^{\mu}\Sigma)^{\dagger} \right ]+\frac{f^{2}}{4}\Tr\left[\chi\Sigma^{\dagger}+\Sigma\chi^{\dagger} \right ]\ ,
\end{split}
\end{equation}
where $f$ is the bare pion decay constant, $F_{\mu\nu}\equiv\partial_{\mu}A^{\rm ext}_{\nu}-\partial_{\nu}A^{\rm ext}_{\mu}$ is the electromagnetic tensor, the covariant derivative is defined as $\nabla_{\mu}\Sigma=\partial_{\mu}\Sigma-ieA^{\rm ext}_{\mu}[Q,\Sigma]$ with $Q={\rm diag}\left(+\tfrac{2}{3},-\tfrac{1}{3},-\tfrac{1}{3}\right)$ being the quark charge matrix and $\chi$ is the scalar-pseudoscalar source which in the presence of a $\theta$-vacuum is $\chi=2B_{0}e^{-i\theta/3}M$. The quark mass matrix, $M$, is
\begin{equation}
\begin{split}
M&=\textrm{diag}{(m_{u},m_{d},m_{s})}=\tfrac{1}{3}(m_{u}+m_{d}+m_{s})\mathbb{1}+\tfrac{1}{2}(m_{u}-m_{d})\lambda_{3}+\tfrac{1}{2\sqrt{3}}(m_{u}+m_{d}-2m_{s})\lambda_{8}\ ,
\end{split}
\end{equation}
where $\lambda_{a}$ are the Gell-Mann matrices. Since $\chi$ has components in the $\mathbb{1}$, $\lambda_{3}$ and $\lambda_{8}$ directions, it is natural to anticipate that the ground state orients itself in these three directions. The most general ansatz that takes this into account is
\begin{equation}
\begin{split}
\Sigma_{\alpha_{i}}={\rm diag}\left (e^{-i\alpha_{1}},e^{-i\alpha_{2}},e^{-i\alpha_{3}} \right)
\end{split}
\end{equation}
with $\sum\alpha_{i}=0$, which ensures that $\det\Sigma_{\alpha_{i}}=1$. Then the tree-level Lagrangian (excluding the uniform magnetic field) is
\begin{equation}
\begin{split}
\label{eq:Ftree}
\mathcal{L}_{2,\rm tree}&=-f^{2}B_{0}\left[m_{u}\cos\phi_{u}(\theta)+m_{d}\cos\phi_{d}(\theta)+m_{s}\cos\phi_{s}(\theta)\right]\ ,
\end{split}
\end{equation}
where $\phi_{u}(\theta)=\frac{\theta}{3}-\alpha_{1}(\theta)$, $\phi_{d}(\theta)=\frac{\theta}{3}-\alpha_{2}(\theta)$ and $\phi_{s}(\theta)=\frac{\theta}{3}-\alpha_{3}(\theta)$. Maximizing the Lagrangian with respect to $\alpha_{i}$ after explicitly imposing the constraint $\sum\alpha_{i}=0$ using a Lagrange multiplier, we get for small values of $\theta$~\cite{Mao:2009sy}, 
\begin{equation}
\begin{split}
\label{eq:treelevelphi}
\phi_{q_{f}}(\theta)&=\frac{\bar{m}}{m_{q_{f}}}\theta+\frac{\bar{m}}{m_{q_{f}}}\left(\frac{\bar{m}^{2}}{m_{q_{f}}^{2}}-\frac{\bar{m}^{3}}{m^{[3]}}\right)\frac{\theta^{3}}{6}+\mathcal{O}(\theta^{5})\ ,
\end{split}
\end{equation}
where $q_{f}$ equals $u$, $d$ or $s$ and higher order corrections can be calculated recursively~\cite{Lu_2020}. For the purposes of calculating the $\mathcal{O}(p^{4})$ free energy near $\theta=0$, we only require the tree level ground state up to cubic order in $\theta$~\cite{Guo:2015oxa}. The reduced mass, $\bar{m}$, which first appears in the linear term and $m^{[3]}$, which appears in the cubic term, are defined as
\begin{align}
\label{eq:mbarm3}
\bar{m}=\left(\frac{1}{m_{u}}+\frac{1}{m_{d}}+\frac{1}{m_{s}}\right)^{-1}\ ,\ \ m^{[3]}=\left(\frac{1}{m_{u}^{3}}+\frac{1}{m_{d}^{3}}+\frac{1}{m_{s}^{3}}\right)^{-1}\ .
\end{align}
We proceed to calculate the one-loop effective potential by parameterizing the field fluctuations around the ground state as
\begin{equation}
\begin{split}
\Sigma=\mathcal{A}_{\alpha_{i}}\exp\left(\tfrac{i\phi_{a}\lambda_{a}}{f}\right)\mathcal{A}_{\alpha_{i}}\ ,
\end{split}
\end{equation}
where $\mathcal{A}_{\alpha_{i}}=\Sigma_{\alpha_{i}}^{1/2}$, $\lambda_{a}$ are the Gell-Mann matrices and there is an implied sum over $a=1,2\dots 8$. We get the following linear and quadratic contributions to the Lagrangian after setting $A_{\mu}^{\rm ext}=0$,
\begin{align}
\nonumber
\mathcal{L}_{2,\rm linear}=&-fB_{0}[m_{u}\sin\phi_{u}(\theta)-m_{d}\sin\phi_{d}(\theta)]\phi_{3}\\
&-\frac{fB_{0}}{\sqrt{3}}[m_{u}\sin\phi_{u}(\theta)+m_{d}\sin\phi_{d}(\theta)-2m_{s}\sin\phi_{s}(\theta)]\phi_{8}\\
\mathcal{L}_{2,\rm quad}=&\frac{1}{2}\partial_{\mu}\phi_{a}\partial^{\mu}\phi_{a}-\frac{1}{2}m_{a}^{2}(\theta)\phi_{a}^{2}+\frac{1}{2}m_{38}^{2}(\theta)\phi_{3}\phi_{8}+\frac{1}{2}m_{38}^{2}(\theta)\phi_{8}\phi_{3}\;,
\end{align}
where the $\theta$-dependent masses $m_{a}(\theta)$ and $m_{38}(\theta)$ are
\begin{align}
m_{1}^{2}(\theta)&=B_{0}[m_{u}\cos\phi_{u}(\theta)+m_{d}\cos\phi_{d}(\theta)]=m_{2}^{2}(\theta)=m_{3}^{2}(\theta)\\
m_{4}^{2}(\theta)&=B_{0}[m_{u}\cos\phi_{u}(\theta)+m_{s}\cos\phi_{s}(\theta)]=m_{5}^{2}(\theta)\\
m_{6}^{2}(\theta)&=B_{0}[m_{d}\cos\phi_{d}(\theta)+m_{s}\cos\phi_{s}(\theta)]=m_{7}^{2}(\theta)\\
m_{8}^{2}(\theta)&=\frac{B_{0}}{3}[m_{u}\cos\phi_{u}(\theta)+m_{d}\cos\phi_{d}(\theta)+4m_{s}\cos\phi_{s}(\theta)]\\
m_{38}^{2}(\theta)&=\frac{B_{0}}{\sqrt{3}}[m_{d}\cos\phi_{d}(\theta)-m_{u}\cos\phi_{u}(\theta)]
\end{align}
In the isospin limit, $m_{38}(\theta)=0$ and the neutral pion and eta do not mix even in the presence of the vacuum angle, $\theta$. We proceed by switching to charge eigenstates, which is physically more transparent in the presence of an external magnetic field, which we now introduce.  Then using the standard definition of the charged eigenstates,
\begin{equation}
\begin{split}
\phi_{a}\lambda_{a}&=
\begin{pmatrix}
\pi^{0}+\tfrac{1}{\sqrt{3}}\eta&\sqrt{2}\pi^{+}&\sqrt{2}K^{+}\\
\sqrt{2}\pi^{-}&-\pi^{0}+\tfrac{1}{\sqrt{3}}\eta&\sqrt{2}K^{0}\\
\sqrt{2}K^{-}&\sqrt{2}\bar{K}^{0}&-\tfrac{2}{\sqrt{3}}\eta
\end{pmatrix}\ ,
\end{split}
\end{equation}
we get the following quadratic Lagrangian
\begin{align}
\nonumber
\mathcal{L}_{2,\rm quad}=&-\frac{1}{2}H^{2}+D_{\mu}\pi^{+}D^{\mu}\pi^{-}-\mathring{m}_{\pi^{\pm}}^{2}(\theta)\pi^{+}\pi^{-}+D_{\mu}K^{+}D^{\mu}K^{-}-\mathring{m}_{K^{\pm}}^{2}(\theta)K^{+}K^{-}+\partial_{\mu}K^{0}\partial^{\mu}\bar{K}^{0}\\
&-\mathring{m}_{K^{0}}^{2}(\theta)K^{0}\bar{K}^{0}+\frac{1}{2}\partial_{\mu}\tilde{\pi}^{0}\partial^{\mu}\tilde{\pi}^{0}-\frac{1}{2}\mathring{m}_{\tilde{\pi}^{0}}^{2}(\theta)\tilde{\pi}^{0}\tilde{\pi}^{0}+\frac{1}{2}\partial_{\mu}\tilde{\eta}\partial^{\mu}\tilde{\eta}-\frac{1}{2}\mathring{m}_{\tilde{\eta}}^{2}(\theta)\tilde{\eta}\tilde{\eta}\ ,
\end{align}
where the covariant derivatives are defined as $D_{\mu}c^{\pm}=(\partial_{\mu}+ieA^{\rm ext}_{\mu})c^{\pm}$ with $c=\pi$ or $K$ and $\mathring{m}_{i}$ indicates the tree level mass of meson $i$. The $\theta$-dependent masses of the charged pions and kaons are
\begin{align}
\label{eq:mp}
\mathring{m}_{\pi^{\pm}}^{2}(\theta)&=B_{0}[m_{u}\cos\phi_{u}(\theta)+m_{d}\cos\phi_{d}(\theta)]\ ,\\
\label{eq:mkc}
\mathring{m}_{K^{\pm}}^{2}(\theta)&=B_{0}[m_{u}\cos\phi_{u}(\theta)+m_{s}\cos\phi_{s}(\theta)]\ ,\\
\label{eq:mk0}
\mathring{m}_{K^{0}}^{2}(\theta)&=B_{0}[m_{d}\cos\phi_{d}(\theta)+m_{s}\cos\phi_{s}(\theta)]=\mathring{m}_{\bar{K}^{0}}^{2}(\theta)\ .
\end{align}
Away from the isospin limit, $\pi^{0}$ and $\eta$ are no longer mass eigenstates. The new mass eigenstates are denoted $\tilde{\pi}^{0}$ and $\tilde{\eta}$ with the following masses
\begin{align}
\label{eq:mp0t}
\mathring{m}_{\tilde{\pi}^{0}}^{2}(\theta)&=m_{3}^{2}(\theta)\cos^{2}\epsilon(\theta)+m_{8}^{2}(\theta)\sin^{2}\epsilon(\theta)-m_{38}^{2}(\theta)\sin2\epsilon(\theta)\ ,\\
\label{eq:metat}
\mathring{m}_{\tilde{\eta}}^{2}(\theta)&=m_{8}^{2}(\theta)\cos^{2}\epsilon(\theta)+m_{3}^{2}(\theta)\sin^{2}\epsilon(\theta)+m_{38}^{2}(\theta)\sin2\epsilon(\theta)\ ,
\end{align}
where $\epsilon(\theta)$ is the tree-level mixing angle between $\pi^{0}$ and $\eta$ for arbitrary values of $\theta$,
\begin{align}
\label{eq:epsilon}
\tan2\epsilon(\theta)=\frac{2m_{38}^{2}(\theta)}{m_{8}^{2}(\theta)-m_{3}^{2}(\theta)}\overset{\theta=0}{=}\frac{\sqrt{3}}{2}\frac{(m_{d}-m_{u})}{m_{s}-\tfrac{m_{u}+m_{d}}{2}}\ .
\end{align}
which depends on the vacuum angle and reduces to the standard $\theta=0$ result~\cite{Gasser:1984gg}. For a next-to-leading order calculation, we require the following one-loop contribution of a pair of charged mesons, which is given by 
\begin{equation}
\begin{split}
I_{H}(m)&=\frac{eH}{2\pi}\sum_{k=0}^{\infty}\int_{p_{0},p_{z}}\ln[p_{0}^{2}+p_{z}^{2}+m_{H}^{2}]\\
\end{split}
\end{equation}
where $m_{H}^{2}=m^{2}+(2k+1)|eH|$ and $\int_{p_{0}p_{z}}\equiv\int\frac{dp_{0}}{2\pi}\frac{dp_{z}}{2\pi}$. The one-loop contribution for each neutral meson is $\frac{1}{2}I_{0}(m)$. In Schwinger proper time form, the integral is
\begin{equation}
\begin{split}
I_{H}(m)&=-\frac{\mu^{2\epsilon}}{(4\pi)^{2}}\int_{0}^{\infty} ds\ \frac{1}{s^{3-\epsilon}}e^{-m^{2}s}\left[\frac{eHs}{\sinh eHs}\right]\ ,
\end{split}
\end{equation}
where $\mu=\sqrt{e^{\gamma_{E}}\Lambda^{2}}$. The integral can be regulated in dimensional regularization. We get two UV divergent pieces, one proportional to $m^{4}$ and the other to $(eH)^{2}$, and a finite piece consistent with Ref.~\cite{Schwinger:1951nm},
\begin{align}
I_{H}(m)=&I_{H}^{\rm div}(m)+I_{H}^{\rm fin}(m)\\
I_{H}^{\rm div}(m)=&-\frac{m^{4}}{2(4\pi)^{2}}\left[\frac{1}{\epsilon}+\frac{3}{2}+\log\frac{\Lambda^{2}}{m^{2}} \right]+\frac{(eH)^{2}}{6(4\pi)^{2}}\left[\frac{1}{\epsilon}+\log\frac{\Lambda^{2}}{m^{2}}\right]\\
I_{H}^{\rm fin}(m)=&-\frac{1}{(4\pi)^{2}}\int_{0}^{\infty} \frac{ds}{s^{3}}e^{-m^{2}s}\left[\frac{eHs}{\sinh eHs}-1+\frac{(eHs)^{2}}{6}\right]\ .
\end{align}
The divergences of the one-loop contributions are canceled by those from the tree level contribution of the $\mathcal{O}(p^{4})$ Lagrangian, which is
\begin{equation}
\begin{split}
\label{eq:L4}
\mathcal{L}_{4}&=L_{6}\left [\Tr(\chi\Sigma^{\dagger}+\chi^{\dagger}\Sigma)\right ]^{2}+L_{7}\left [\Tr(\chi\Sigma^{\dagger}-\chi^{\dagger}\Sigma)\right ]^{2}+L_{8}\Tr \left(\Sigma \chi^{\dagger}\Sigma \chi^{\dagger}+\chi\Sigma^{\dagger}\chi\Sigma^{\dagger} \right )\\
&+L_{10}\Tr\left [\Sigma F_{\mu\nu}^{L}\Sigma^{\dagger}F^{R\mu\nu}\right ]+H_{1}\Tr\left [F_{\mu\nu}^{R}F^{R\mu\nu}+F_{\mu\nu}^{L}F^{L\mu\nu} \right ]+H_{2}\Tr(\chi \chi^{\dagger})\ ,
\end{split}
\end{equation}
where $F_{\mu\nu}^{R}=F_{\mu\nu}^{L}=-eQF_{\mu\nu}$, where $L_{i}$ and $H_{i}$ are the relevant low and high energy constants encoding quark physics defined as
\begin{align}
\label{lihi}
L_{i}&=L_{i}^{r}+\Gamma_{i}\lambda,\ H_{i}=H_{i}^{r}+\Delta_{i}\lambda\ ,\ \lambda=-\frac{\Lambda^{-2\epsilon}}{2(4\pi)^{2}}\left(\frac{1}{\epsilon}+1\right)\ ,
\end{align}
where we require the following $\Gamma_{i}$ and $\Delta_{i}$ for renormalization,
\begin{equation}
\begin{split}
\Gamma_{6}&=\frac{11}{144},\ \Gamma_{7}=0,\ \Gamma_{8}=\frac{5}{48},\ \Gamma_{10}=-\frac{1}{4},\ \Delta_{1}=-\frac{1}{8},\ \Delta_{2}=\frac{5}{24}\ .
\end{split}
\end{equation} 
The divergences of the one-loop piece in $\mathcal{F}_{1}$ below is canceled by the contributions from $\mathcal{F}_{\rm ct}$
\begin{align}
\mathcal{F}_{1}=&I_{H}[\mathring{m}_{\pi^{\pm}}(\theta)]+I_{H}[\mathring{m}_{K^{\pm}}(\theta)]+I_{0}[\mathring{m}_{K^{0}}(\theta)]+\frac{1}{2}I_{0}[\mathring{m}_{\tilde{\pi}^{0}}(\theta)]+\frac{1}{2}I_{0}[\mathring{m}_{\tilde{\eta}}(\theta)]\\\nonumber
\mathcal{F}_{\rm ct}=&-16L_{6}B_{0}^{2}[m_{u}\cos\phi_{u}(\theta)+m_{d}\cos\phi_{d}(\theta)+m_{s}\cos\phi_{s}(\theta)]^{2}\\\nonumber
&+16L_{7}B_{0}[m_{u}\sin\phi_{u}(\theta)+m_{d}\sin\phi_{d}(\theta)+m_{s}\sin\phi_{s}(\theta)]^{2}\\\nonumber
&-8L_{8}B_{0}^{2}\left[m_{u}^{2}\cos(2\phi_{u}(\theta))+m_{d}^{2}\cos(2\phi_{d}(\theta))+m_{s}^{2}\cos(2\phi_{s}(\theta))\right]\\
&-4H_{2}B_{0}^{2}(m_{u}^{2}+m_{d}^{2}+m_{s}^{2})-\frac{4}{3}(L_{10}+2H_{1})(eH)^{2}\ 
\end{align}
Then combining these with $\mathcal{F}_{\rm tree}$ in Eq.~(\ref{eq:Ftree}), we get the full free energy $\mathcal{F}$ is
\begin{align}
\mathcal{F}=\mathcal{F}_{0}+\mathcal{F}_{H}\ ,
\end{align}
where the $H$-independent contribution, $\mathcal{F}_{0}$, is
\begin{equation}
\begin{split}
\label{eq:F0}
\mathcal{F}_{0}=&-f^{2}B_{0}[m_{u}\cos\phi_{u}(\theta)+m_{d}\cos\phi_{d}(\theta)+m_{s}\cos\phi_{s}(\theta)]\\
&-\frac{2\mathring{m}^{4}_{\pi^{\pm}}(\theta)}{4(4\pi)^{2}}\left [\frac{1}{2}+\log\left (\frac{\Lambda^{2}}{\mathring{m}_{\pi^{\pm}}^{2}(\theta)} \right ) \right ]-\frac{2\mathring{m}_{K^{\pm}}^{4}(\theta)}{4(4\pi)^{2}}\left [\frac{1}{2}+\log\left (\frac{\Lambda^{2}}{\mathring{m}_{K^{\pm}}^{2}(\theta)} \right ) \right ]\\
&-\frac{2\mathring{m}_{K^{0}}^{4}(\theta)}{4(4\pi)^{2}}\left [\frac{1}{2}+\log\left (\frac{\Lambda^{2}}{\mathring{m}_{K^{0}}^{2}(\theta)} \right ) \right ]\\
&-\frac{\mathring{m}_{\tilde{\pi}^{0}}^{4}(\theta)}{4(4\pi)^{2}}\left [\frac{1}{2}+\log\left (\frac{\Lambda^{2}}{\mathring{m}_{\tilde{\pi}^{0}}^{2}(\theta)} \right ) \right ]-\frac{\mathring{m}_{\tilde{\eta}}^{4}(\theta)}{4(4\pi)^{2}}\left [\frac{1}{2}+\log\left (\frac{\Lambda^{2}}{\mathring{m}_{\tilde{\eta}}^{2}(\theta)} \right ) \right ]\\
&-16L^{r}_{6}B_{0}^{2}[m_{u}\cos\phi_{u}(\theta)+m_{d}\cos\phi_{d}(\theta)+m_{s}\cos\phi_{s}(\theta)]^{2}\\
&+16L^{r}_{7}B_{0}^{2}[m_{u}\sin\phi_{u}(\theta)+m_{d}\sin\phi_{d}(\theta)+m_{s}\sin\phi_{s}(\theta)]^{2}\\
&-8L^{r}_{8}B_{0}^{2}\left[m_{u}^{2}\cos(2\phi_{u}(\theta))+m_{d}^{2}\cos(2\phi_{d}(\theta))+m_{s}^{2}\cos(2\phi_{s}(\theta))\right]\\
&-4H^{r}_{2}B_{0}^{2}\left[m_{u}^{2}+m_{d}^{2}+m_{s}^{2}\right]\ ,
\end{split}
\end{equation}
and the $H$-dependent contribution, $\mathcal{F}_{H}$, is
\begin{equation}
\begin{split}
\label{eq:FH}
\mathcal{F}_{H}(\theta)&=\frac{1}{2}H_{R}^{2}+\frac{(eH)^{2}}{(4\pi)^{2}}\left[\mathfrak{I}_{H}(\tfrac{\mathring{m}_{\pi^{\pm}}^{2}(\theta)}{eH})+\mathfrak{I}_{H}(\tfrac{\mathring{m}_{K^{\pm}}^{2}(\theta)}{eH})\right]\ ,
\end{split}
\end{equation}
where the charged pion and kaon masses are defined in Eqs.~(\ref{eq:mp}) and (\ref{eq:mkc}), $\mathfrak{I}_{H}$ is the Schwinger integral defined in Eq.~(\ref{eq:Schwinger-integral}) and $H_{R}$ is the renormalized magnetic field,
\begin{equation}
\begin{split}
H_{R}&=Z_{H}H\ ,\ Z_{H}=1-\frac{4e^{2}}{3}\left [L^{r}_{10}+2H^{r}_{1}-\frac{1}{4(4\pi)^{2}}\left\{\log\frac{\Lambda^{2}}{\mathring{m}_{\pi^{\pm}}^{2}(\theta)}+\log\frac{\Lambda^{2}}{\mathring{m}_{K^{\pm}}^{2}(\theta)}-2\right\}\right ]\ ,
\end{split}
\end{equation}
where a corresponding charge renormalization, i.e. $e_{R}=Z_{H}^{-1}e$ ensures that $eH=e_{R}H_{R}$ remains unaltered and finite.
\subsection{Topological Susceptibility and Fourth Cumulant}
\label{sec:top}
\noindent
Using the $\phi_{q_{f}}$-dependent free energy, we can calculate the topological susceptibility and the fourth cumulant, which are defined as
\begin{equation}
\begin{split}
\label{eq:chitc40}
\chi_{t}=\left.\frac{\partial^{2}\mathcal{F}(\theta,H)}{\partial \theta^{2}}\right|_{\theta=0}\ ,\ c_{4}=\left.\frac{\partial^{4}\mathcal{F}(\theta,H)}{\partial \theta^{4}}\right|_{\theta=0}\ .
\end{split}
\end{equation}
At $\mathcal{O}(p^{4})$, we can use the tree level values of $\phi_{q_{f}}$ in Eq.~$(\ref{eq:treelevelphi})$. We find that the topological susceptibility is
\begin{equation}
\begin{split}
\label{eq:chit}
&\chi_{t}=\bar{m}B_{0}f^{2}+32B_{0}^{2}\bar{m}(m_{u}+m_{d}+m_{s})L_{6}^{r}+96B_{0}^{2}\bar{m}^{2}(3L^{r}_{7}+L^{r}_{8})\\
&+\frac{B_{0}^{2}\bar{m}^{2}}{(4\pi)^{2}}\sum_{(ij)}\left(\frac{1}{m_{i}}+\frac{1}{m_{j}}\right)(m_{i}+m_{j})\log\frac{\Lambda^{2}}{B_{0}(m_{i}+m_{j})}\\
&+\frac{\bar{m}^{2}B_{0}}{2(4\pi)^{2}}\left[\left(\frac{1}{m_{u}}+\frac{1}{m_{d}}\right)+\frac{2\sin\epsilon\cos\epsilon}{\sqrt{3}}\frac{m_{d}-m_{u}}{m_{u}m_{d}}+\frac{2\sin^{2}\epsilon}{3}\left(\frac{2}{m_{s}}-\frac{m_{u}+m_{d}}{m_{u}m_{d}}\right)\right]\mathring{m}_{\tilde{\pi}^{0}}^{2}\log\frac{\Lambda^{2}}{\mathring{m}_{\tilde{\pi}^{0}}^{2}}\\
&+\frac{\bar{m}^{2}B_{0}}{2(4\pi)^{2}}\left[\frac{1}{3}\left(\frac{1}{m_{u}}+\frac{1}{m_{d}}+\frac{4}{m_{s}}\right)-\frac{2\sin\epsilon\cos\epsilon}{\sqrt{3}}\frac{m_{d}-m_{u}}{m_{u}m_{d}}-\frac{2\sin^{2}\epsilon}{3}\left(\frac{2}{m_{s}}-\frac{m_{u}+m_{d}}{m_{u}m_{d}}\right)\right]\\
&\times\mathring{m}_{\tilde{\eta}}^{2}\log\frac{\Lambda^{2}}{\mathring{m}_{\tilde{\eta}}^{2}}-\frac{B_{0}\bar{m}^{2}(eH)}{(4\pi)^{2}} \left[\left(\frac{1}{m_{u}}+\frac{1}{m_{d}}\right) \mathcal{I}_{H,2}(\tfrac{\mathring{m}_{\pi^{\pm}}^{2}}{eH})+\left(\frac{1}{m_{u}}+\frac{1}{m_{s}}\right)\mathcal{I}_{H,2}(\tfrac{\mathring{m}_{K^{\pm}}^{2}}{eH})\right]\ ,
\end{split}
\end{equation}
where $\bar{m}$ is the reduced mass defined in Eq.~(\ref{eq:mbarm3}), the sum over $(ij)$ represents cyclic permutations over the quark flavors, i.e. $ij=ud,ds,su$ and the integrals $\mathcal{I}_{H,2}$ are defined in Eq.~(\ref{eq:IHn}) with the explicit analytic form given in Eq.~(\ref{eq:IH2}). $\epsilon$ is the mixing angle between $\pi^{0}$ and $\eta$, which is defined in Eq. (\ref{eq:epsilon}), in the $\theta=0$ vacuum, i.e. $\epsilon(0)\equiv\epsilon$. For the meson masses $\mathring{m}_{i}\equiv\mathring{m}_{i}(0)$. The $H$-independent contribution to the topological susceptibility is consistent with Ref.~\cite{Bernard:2012fw}. Similarly, we find that the fourth cumulant is
\begin{equation}
\begin{split}
\label{eq:c4}
c_{4}=&-\frac{B_{0}\bar{m}^{4}f^{2}}{m^{[3]}}-32B_{0}^{2}\bar{m}^{4}L^{r}_{6}\left[6\sum_{(ij)}\frac{1}{m_{i}m_{j}}+4\sum_{q}\frac{1}{m_{q}^{2}}+\sum_{(ijk)}m_{i}\left(\frac{1}{m_{j}^{3}}+\frac{1}{m_{k}^{3}}\right)\right]\\
&-384B_{0}^{2}\frac{\bar{m}^{5}}{m^{[3]}}(3L^{r}_{7}+L^{r}_{8})+\frac{6B_{0}^{2}\bar{m}^{5}}{(4\pi)^{2}}\left[\frac{1}{m^{[3]}}+2\sum_{(ij)}\frac{1}{m_{i}m_{j}}\left(\frac{1}{m_{i}}+\frac{1}{m_{j}}\right)+\frac{3}{m_{u}m_{d}m_{s}}\right]\\
&-\frac{B_{0}^{2}\bar{m}^{5}}{(4\pi)^{2}}\left[\sum_{(ijk)}(m_{i}+m_{j})^{2}\left\{\frac{1}{m_{i}m_{j}}\left(\frac{1}{m_{i}^{3}}+\frac{1}{m_{j}^{3}}+\frac{4}{m_{k}^{3}}\right)+\frac{3}{m_{i}^{2}m_{j}^{2}}\left(\frac{1}{m_{i}}+\frac{1}{m_{j}}\right)\right.\right.\\
&\left.\left.-\frac{3}{m_{i}m_{j}m_{k}}\left(\frac{1}{m_{i}^{2}}+\frac{1}{m_{j}^{2}}\right)+\frac{6}{m_{i}^{2}m_{j}^{2}m_{k}}\right\}\log\frac{\Lambda^{2}}{B_{0}(m_{i}+m_{j})}\right ]+c_{4,\tilde{\pi}^{0}}+c_{4,\tilde{\eta}}\\
&+\frac{B_{0}\bar{m}^{4}(eH)}{(4\pi)^{2}}\left[\left(\frac{1}{m_{u}^{3}}+\frac{1}{m_{d}^{3}}\right)\mathcal{I}_{H,2}(\tfrac{\mathring{m}_{\pi^{\pm}}^{2}}{eH})+\left(\frac{1}{m_{u}^{3}}+\frac{1}{m_{s}^{3}}\right)\mathcal{I}_{H,2}(\tfrac{\mathring{m}_{K^{\pm}}^{2}}{eH})\right]\\
&-\frac{3B_{0}^{2}\bar{m}^{5}}{(4\pi)^{2}}\left[\frac{1}{\bar{m}_{ud}}\left(\frac{1}{m_{u}}+\frac{1}{m_{d}}\right)^{2}\mathcal{I}_{H,1}(\tfrac{\mathring{m}_{\pi^{\pm}}^{2}}{eH})+\frac{1}{\bar{m}_{us}}\left(\frac{1}{m_{u}}+\frac{1}{m_{s}}\right)^{2}\mathcal{I}_{H,1}(\tfrac{\mathring{m}_{K^{\pm}}^{2}}{eH})\right]\ ,
\end{split}
\end{equation}
where $\bar{m}_{ud}$ and $\bar{m}_{us}$ are defined as
\begin{align}
\label{eq:mbud}
\bar{m}_{ud}^{-1}&=\left[\frac{1}{m_{u}}+\frac{1}{m_{d}}-\frac{3}{m_{s}}+\frac{4}{m_{s}^{3}}\left(\frac{1}{m_{u}^{2}}+\frac{1}{m_{d}^{2}}-\frac{1}{m_{u}m_{d}}\right)^{-1} \right]\\
\label{eq:mbus}
\bar{m}_{us}^{-1}&=\left[\frac{1}{m_{u}}+\frac{1}{m_{s}}-\frac{3}{m_{d}}+\frac{4}{m_{d}^{3}}\left(\frac{1}{m_{u}^{2}}+\frac{1}{m_{s}^{2}}-\frac{1}{m_{u}m_{s}}\right)^{-1} \right]\ ,
\end{align}
$m^{[3]}$ is defined in Eq.~(\ref{eq:mbarm3}), $(ijk)$ represents a cyclic sum over quark flavors, i.e. $ijk=uds,dsu,sud$ and the integrals $\mathcal{I}_{H,n}$ are defined in Eqs.~(\ref{eq:IHn}), (\ref{eq:IH2}) and (\ref{eq:IH1}). The contribution from the neutral pion and eta are denoted $c_{4,\tilde{\pi}^{0}}$ and $c_{4,\tilde{\eta}}$ respectively. They are formally the fourth derivative of the following contributions
\begin{equation}
\tilde{I}(\mathring{m}(\theta))=-\frac{\mathring{m}^{4}(\theta)}{4(4\pi)^{2}}\left [\frac{1}{2}+\log\left (\frac{\Lambda^{2}}{\mathring{m}^{2}(\theta)} \right ) \right ]\ ,
\end{equation}
where $\mathring{m}=\mathring{m}_{\tilde{\pi}^{0}},\mathring{m}_{\tilde{\eta}}$ for the neutral pion and the eta respectively. Due to the cumbersome nature of the full result~\cite{Bernard:2012fw,GomezNicola:2019myi}, we only state the result valid in the isospin limit, which is
\begin{align}
c_{4,\tilde{\pi}^{0}}&=\left.\frac{\partial I[\mathring{m}_{\tilde{\pi}^{0}}(\theta)]}{\partial\theta^{4}}\right |_{\theta=0}\overset{m_{u}=m_{d}}{=}\frac{6B_{0}^{2}\bar{m}^{5}}{(4\pi)^{2}m_{s}\hat{m}^{2}}\left[1+2\frac{m_{s}}{\hat{m}}-\frac{4}{3}\left(\frac{\hat{m}^{2}}{m_{s}^{2}}+2\frac{m_{s}}{\hat{m}}\right)\log\frac{\Lambda^{2}}{2B_{0}\hat{m}}\right]\\
\nonumber
c_{4,\tilde{\eta}}&=\left.\frac{\partial I[\mathring{m}_{\tilde{\eta}}(\theta)]}{\partial\theta^{4}}\right |_{\theta=0}\overset{m_{u}=m_{d}}{=}\frac{6B_{0}^{2}\bar{m}^{5}}{(4\pi)^{2}m_{s}\hat{m}^{2}}\left[1+\frac{2m_{s}}{9\hat{m}}+\frac{4\hat{m}}{3m_{s}}+\frac{4\hat{m}^{2}}{9m_{s}^{2}}\right.\\
&\left.-\frac{2}{27}\left(20+\frac{10\hat{m}}{m_{s}}+\frac{17m_{s}}{\hat{m}}+\frac{4\hat{m}^{2}}{m_{s}^{2}}+\frac{2m_{s}^{2}}{\hat{m}^{2}}+\frac{\hat{m}^{3}}{m_{s}^{3}}\right)\log\frac{\Lambda^{2}}{\tfrac{2B_{0}}{3}(\hat{m}+2m_{s})}\right]\ ,
\end{align}
where $\hat{m}=\frac{m_{u}+m_{d}}{2}$ is the average light quark mass.
\subsection{Chiral Condensate and Chiral Susceptibility}
\label{sec:cc}
\noindent
The chiral condensate and the chiral susceptibilities are defined as 
\begin{equation}
\begin{split}
\langle\bar{q}q\rangle=\left.\frac{\partial \mathcal{F}(\theta,H)}{\partial m_{q}}\right|_{\theta=0}\ ,\ \chi_{q}=\left.\frac{\partial^{2} \mathcal{F}(\theta,H)}{\partial m_{q}^{2}}\right|_{\theta=0} \ ,
\end{split}
\end{equation}
where we use the $\theta=0$ values for the free energy. The resulting expressions for the up quark condensate is
\begin{equation}
\begin{split}
\label{eq:uu}
\langle\bar{u}u\rangle&=-f^{2}B_{0}-32B_{0}^{2}(m_{u}+m_{d}+m_{s})L^{r}_{6}-8B_{0}^{2}m_{u}(2L^{r}_{8}+H^{r}_{2})\\
&-\frac{B_{0}^{2}(m_{u}+m_{d})}{(4\pi)^{2}}\log\frac{\Lambda^{2}}{B_{0}(m_{u}+m_{d})}-\frac{B_{0}^{2}(m_{u}+m_{s})}{(4\pi)^{2}}\log\frac{\Lambda^{2}}{B_{0}(m_{u}+m_{s})}\\
&-\frac{B_{0}}{2(4\pi)^{2}}\left[\mathring{m}_{\tilde{\pi}^{0}}^{2}\log\frac{\Lambda^{2}}{\mathring{m}_{\tilde{\pi}^{0}}^{2}}(\cos\epsilon+\tfrac{1}{\sqrt{3}}\sin\epsilon)^{2}+\mathring{m}_{\tilde{\eta}}^{2}\log\frac{\Lambda^{2}}{\mathring{m}_{\tilde{\eta}}^{2}}(-\sin\epsilon+\tfrac{1}{\sqrt{3}}\cos\epsilon)^{2}\right]\\
&+\frac{B_{0}(eH)}{(4\pi)^{2}}\left[\mathcal{I}_{H,2}(\tfrac{\mathring{m}_{\pi^{\pm}}^{2}}{eH})+\mathcal{I}_{H,2}(\tfrac{\mathring{m}_{K^{\pm}}^{2}}{eH})\right]\ ,
\end{split}
\end{equation}
the down quark condensate is
\begin{equation}
\begin{split}
\label{eq:dd}
\langle\bar{d}d\rangle&=-f^{2}B_{0}-32B_{0}^{2}(m_{u}+m_{d}+m_{s})L^{r}_{6}-8B_{0}^{2}m_{d}(2L^{r}_{8}+H^{r}_{2})\\
&-\frac{B_{0}^{2}(m_{u}+m_{d})}{(4\pi)^{2}}\log\frac{\Lambda^{2}}{B_{0}(m_{u}+m_{d})}-\frac{B_{0}^{2}(m_{d}+m_{s})}{(4\pi)^{2}}\log\frac{\Lambda^{2}}{B_{0}(m_{d}+m_{s})}\\
&-\frac{B_{0}}{2(4\pi)^{2}}\left[\mathring{m}_{\tilde{\pi}^{0}}^{2}\log\frac{\Lambda^{2}}{\mathring{m}_{\tilde{\pi}^{0}}^{2}}(\cos\epsilon-\tfrac{1}{\sqrt{3}}\sin\epsilon)^{2}+\mathring{m}_{\tilde{\eta}}^{2}\log\frac{\Lambda^{2}}{\mathring{m}_{\tilde{\eta}}^{2}}(\sin\epsilon+\tfrac{1}{\sqrt{3}}\cos\epsilon)^{2}\right]\\
&+\frac{B_{0}(eH)}{(4\pi)^{2}}\mathcal{I}_{H,2}(\tfrac{\mathring{m}_{\pi^{\pm}}^{2}}{eH})\ ,
\end{split}
\end{equation}
and the strange quark condensate is
\begin{equation}
\begin{split}
\label{eq:ss}
\langle\bar{s}s\rangle&=-f^{2}B_{0}-32B_{0}^{2}(m_{u}+m_{d}+m_{s})L^{r}_{6}-8B_{0}^{2}m_{s}(2L^{r}_{8}+H^{r}_{2})\\
&-\frac{B_{0}^{2}(m_{u}+m_{s})}{(4\pi)^{2}}\log\frac{\Lambda^{2}}{B_{0}(m_{u}+m_{s})}-\frac{B_{0}^{2}(m_{d}+m_{s})}{(4\pi)^{2}}\log\frac{\Lambda^{2}}{B_{0}(m_{d}+m_{s})}\\
&-\frac{2B_{0}}{3(4\pi)^{2}}\left[\mathring{m}_{\tilde{\pi}^{0}}^{2}\log\frac{\Lambda^{2}}{\mathring{m}_{\tilde{\pi}^{0}}^{2}}\sin^{2}\epsilon+\mathring{m}_{\tilde{\eta}}^{2}\log\frac{\Lambda^{2}}{\mathring{m}_{\tilde{\eta}}^{2}}\cos^{2}\epsilon\right]+\frac{B_{0}(eH)}{(4\pi)^{2}}\mathcal{I}_{H,2}(\tfrac{\mathring{m}_{K^{\pm}}^{2}}{eH})\ ,
\end{split}
\end{equation}
where the $H=0$ quark condensates are in agreement with those of Ref.~\cite{Gasser:1984gg}. The up quark chiral susceptibility is
\begin{equation}
\begin{split}
\label{eq:chiu}
\chi_{u}=&-8B_{0}^{2}(4L^{r}_{6}+2L^{r}_{8}+H^{r}_{2})-\frac{B_{0}^{2}}{(4\pi)^{2}}\left[-2+\log\frac{\Lambda^{2}}{B_{0}(m_{u}+m_{d})}+\log\frac{\Lambda^{2}}{B_{0}(m_{u}+m_{s})}\right]\\
&-\frac{B_{0}^{2}}{2(4\pi)^{2}}\left(-1+\log\frac{\Lambda^{2}}{\mathring{m}_{\tilde{\pi}^{0}}^{2}}\right)(\cos\epsilon+\tfrac{1}{\sqrt{3}}\sin\epsilon)^{4}\\
&-\frac{B_{0}^{2}}{2(4\pi)^{2}}\left(-1+\log\frac{\Lambda^{2}}{\mathring{m}_{\tilde{\eta}}^{2}}\right)(-\sin\epsilon+\tfrac{1}{\sqrt{3}}\cos\epsilon)^{4}\\
&-\frac{B_{0}^{2}}{(4\pi)^{2}}\left[\mathcal{I}_{H,1}(\tfrac{\mathring{m}^{2}_{\pi^{\pm}}}{eH})+\mathcal{I}_{H,1}(\tfrac{\mathring{m}^{2}_{K^{\pm}}}{eH})\right]\ ,
\end{split}
\end{equation}
the down quark susceptibility is
\begin{equation}
\begin{split}
\label{eq:chid}
\chi_{d}=&-8B_{0}^{2}(4L^{r}_{6}+2L^{r}_{8}+H^{r}_{2})-\frac{B_{0}^{2}}{(4\pi)^{2}}\left[-2+\log\frac{\Lambda^{2}}{B_{0}(m_{u}+m_{d})}+\log\frac{\Lambda^{2}}{B_{0}(m_{d}+m_{s})}\right]\\
&-\frac{B_{0}^{2}}{2(4\pi)^{2}}\left(-1+\log\frac{\Lambda^{2}}{\mathring{m}_{\tilde{\pi}^{0}}^{2}}\right)(\cos\epsilon-\tfrac{1}{\sqrt{3}}\sin\epsilon)^{4}\\
&-\frac{B_{0}^{2}}{2(4\pi)^{2}}\left(-1+\log\frac{\Lambda^{2}}{\mathring{m}_{\tilde{\eta}}^{2}}\right)(\sin\epsilon+\tfrac{1}{\sqrt{3}}\cos\epsilon)^{4}\\
&-\frac{B_{0}^{2}}{(4\pi)^{2}}\mathcal{I}_{H,1}(\tfrac{\mathring{m}^{2}_{\pi^{\pm}}}{eH})\ ,
\end{split}
\end{equation}
and the strange quark susceptibility is
\begin{equation}
\begin{split}
\label{eq:chis}
\chi_{s}=&-8B_{0}^{2}(4L^{r}_{6}+2L^{r}_{8}+H^{r}_{2})-\frac{B_{0}^{2}}{(4\pi)^{2}}\left[-2+\log\frac{\Lambda^{2}}{B_{0}(m_{u}+m_{s})}+\log\frac{\Lambda^{2}}{B_{0}(m_{d}+m_{s})}\right]\\
&-\frac{8B_{0}^{2}}{9(4\pi)^{2}}\left(-1+\log\frac{\Lambda^{2}}{m_{\tilde{\pi}^{0}}^{2}}\right)\sin^{4}\epsilon-\frac{8B_{0}^{2}}{9(4\pi)^{2}}\left(-1+\log\frac{\Lambda^{2}}{m_{\tilde{\eta}}^{2}}\right)\cos^{4}\epsilon\\
&-\frac{B_{0}^{2}}{(4\pi)^{2}}\mathcal{I}_{H,1}(\tfrac{\mathring{m}^{2}_{K^{\pm}}}{eH})\ .
\end{split}
\end{equation}
\subsection{Sum Rules in Three-Flavor $\chi$PT}
\label{sec:sumrules}
In this subsection, we discuss the relationship of the shifts in the topological susceptibility and the fourth cumulant due to the background magnetic field with those of the quark condensates and the quark susceptibilities. We denote the shift of the quantity $\mathcal{O}$ by $\mathcal{O}_{H}$. Firstly, note that the shifts in the three quark condensates are given by the last terms of Eqs.~(\ref{eq:uu}), (\ref{eq:dd}) and (\ref{eq:ss}) and those in the quark susceptibilities are the last terms of Eqs.~(\ref{eq:chiu}), (\ref{eq:chid}) and (\ref{eq:chis}). We get the following sum rules for the shifts of the quark condensates and the quark susceptibilties
\begin{align}
\label{eq:sumrule-qq}
\langle\bar{u}u\rangle_{H}&=\langle\bar{d}d\rangle_{H}+\langle\bar{s}s\rangle_{H}\\
\label{eq:sumrule-chiq}
\chi_{u,H}&=\chi_{d,H}+\chi_{s,H}\ ,
\end{align}
which are generalizations of the sum rule at $H=0$ and the isospin limit first discussed in Ref.~\cite{Gasser:1984gg}. We should emphasize that these sum rules for the shifts due to a background magnetic field hold away from the isospin limit. The origin of the sum rule for the quark condensates and susceptibilities can be explained through the contribution of the charged mesons to the free energy in Eq.~(\ref{eq:FH}), in particular the charged pions and kaons both contribute to the up-quark condensate shift with only the charged pions contribution to the down-quark condensate shift and only the charged kaons contributing to the strange-quark condensate shift. The same line of argument also applies for the quark susceptibilties.

The shift for the topological susceptibility, given by the last term of Eq.~(\ref{eq:chit}), is proportional to the integral $\mathcal{I}_{n,2}(x)$, where the argument $x=\tfrac{\mathring{m}_{\pi^{\pm}}^{2}}{eH}$ or $\tfrac{\mathring{m}_{K^{\pm}}^{2}}{eH}$. Combining the shift of the topological susceptibility with the sum rule 
for the quark condensate shift of Eq.~(\ref{eq:sumrule-qq}), we get
\begin{align}
\label{eq:sumrule-chitH}
\chi_{t,H}=-\bar{m}^{2}\sum_{q_{f}=uds}\frac{\langle\bar{q_{f}}q_{f}\rangle_{H}}{m_{q_{f}}}\ ,
\end{align}
where $\bar{m}$ is the reduced mass defined in Eq.~(\ref{eq:mbarm3}), $q_{f}$ is the quark flavor index, i.e. $q_{f}=u,d,$ or $s$, $m_{q_{f}}$ is the quark mass of flavor $q_{f}$ and $\langle\bar{q}_{f}q_{f}\rangle_{H}$ is the finite $H$ shift in the quark condensate for quark flavor $q$.  Due to magnetic catalysis, $\langle\bar{q}_{f}q_{f}\rangle_{H}$ is negative for finite $H$ and consequently the shift of the topological susceptibility due to a finite magnetic field is positive and vanishes when $H=0$. Similar to the zero magnetic field susceptibility, the shift of the topological susceptibility even at finite $H$ vanishes for zero quark masses since the reduced mass squared goes to zero faster than the $m_{q}$ in the denominator.

Finally, the shift in the fourth cumulant is given by the last two lines of Eq.~(\ref{eq:c4}). It depends not just on $\mathcal{I}_{n,2}(x)$ but also on $\mathcal{I}_{n,1}(x)$, where the argument $x$ depends on the bare pion and kaon mass and the magnetic field as for the topological susceptibility shift. Since the shift of chiral susceptibilities depends on $\mathcal{I}_{n,1}(x)$ and the shift of the quark condensates depends on $\mathcal{I}_{n,2}(x)$, we deduce a sum rule that relates the shift of the fourth cumulants to the shifts of the quark condensates and susceptibilities~\cite{Adhikari:2021lbl},
\begin{align}
\label{eq:sumrule-c4H}
c_{4,H}=&\bar{m}^{4}\sum_{q_{f}=uds}\frac{\langle\bar{q}_{f}q_{f}\rangle_{H}}{m_{q_{f}}^{3}}+3\bar{m}^{5}\left[\frac{1}{\bar{m}_{ud}}\left(\frac{1}{m_{u}}+\frac{1}{m_{d}}\right)^{2}\chi_{d,H}+\frac{1}{\bar{m}_{us}}\left(\frac{1}{m_{u}}+\frac{1}{m_{s}}\right)^{2}\chi_{s,H}\right]\ ,
\end{align}
where the masses $\bar{m}_{ud}$ and $\bar{m}_{us}$ are defined in Eqs.~(\ref{eq:mbud}) and (\ref{eq:mbus}). 
In the large $m_{s}$ limit, the shifts in both topological susceptibility and the fourth cumulant reduced in Eqs. (\ref{eq:sumrule-chitH}) and (\ref{eq:sumrule-c4H}) to the two-flavor results~\cite{Adhikari:2021xra} since the three-flavor reduced mass becomes the two-flavor reduced mass, terms inversely proportional to $m_{s}$ are suppressed including the shift in the strange quark condensate and the strange quark susceptibility. Finally, the mass $\bar{m}_{ud}$ becomes the two-flavor reduced mass.
\subsection{$n$-flavor $\chi$PT} 
\label{sec:nf}
\noindent
Here we generalize the three-flavor calculation to $n$-flavor $\chi$PT with $n_{Q}$ pairs of charged mesons. The relationship between $n_{Q}$ and $n$ depends on the choice of the charge matrix. Using the structure of the three-flavor free energy in Eq.~(\ref{eq:F0}), it is straightforward to deduce that in the $n$-flavor case, the free energy becomes
\begin{equation}
\begin{split}
\mathcal{F}=&\frac{1}{2}H_{R}^{2}-f_{n}^{2}B_{n}\sum_{f=1}^{n}m_{q_{f}}\cos\phi_{q_{f}}-\sum_{i=1}^{n^{2}-1}\frac{\mathring{m}_{i}^{4}(\theta)}{4(4\pi)^{2}}\left[\frac{1}{2}+\log\frac{\Lambda^{2}}{\mathring{m}_{i}^{2}} \right ]-16L^{r}_{6}B^{2}_{n}\Big(\sum_{f=1}^{n}m_{q_{f}}\cos\phi_{q_{f}}\Big)^{2}\\
&+16L^{r}_{7}B_{n}^{2}\Big(\sum_{f=1}^{n}m_{q_{f}}\sin\phi_{q_{f}}\Big)^{2}-8L^{r}_{8}B_{n}^{2}\Big(\sum_{f=1}^{n}m_{q_{f}}^{2}\cos(2\phi_{q_{f}})\Big)-4H^{r}_{2}B_{n}^{2}\Big(\sum_{f=1}^{n}m_{q_{f}}^{2}\Big)\\
&+\sum_{i=1}^{n_{Q}}\frac{(eH)^{2}}{(4\pi)^{2}}\mathfrak{I}_{H}(\tfrac{\mathring{m}_{i}^{2}(\theta)}{eH})\ ,
\end{split}
\end{equation}
where $f_{n}$ is the $n$-flavor pion decay constant, $m_{q_{f}}$ is the mass of the $f^{th}$ quark flavor and $\mathring{m}_{i}$ is the mass of the $i^{th}$ pseudo-Goldstone boson and $\mathfrak{I}_{H}$ is the Schwinger integral defined in Eq.~(\ref{eq:Schwinger-integral}). $H_{R}$ is the renormalized magnetic field, where $H_{R}=Z_{H}H$ and
\begin{equation}
\begin{split}
Z_{H}=1+\frac{e^{2}}{6}\left [h_{Q}+\frac{1}{(4\pi)^{2}}\left\{\sum_{i=1}^{n_{Q}}\ln\frac{\Lambda^{2}}{\mathring{m}_{i}(\theta)}-n_{Q}\right\}\right]\ .
\end{split}
\end{equation}
$h_{Q}$ is a linear function of $2H_{1}^{r}+L^{r}_{10}$ with coefficients that depend on the quark charge matrix. Since $h_{Q}$ is $\theta$-independent, the detailed form is irrelevant to the calculation of the topological susceptibility and the fourth cumulant. Assuming degenerate quarks as in Ref.~\cite{Guo:2015oxa}, $m_{q_{f}}=m_{q}$ and $\mathring{m}_{i}^{2}(\theta)=\mathring{m}^{2}(\theta)=2B_{n}m_{q}\cos\frac{\theta}{n}$, the $n$-flavor $\chi$PT free energy is
\begin{equation}
\begin{split}
\mathcal{F}(\theta)&=\frac{1}{2}H_{R}^{2}-\frac{1}{2}nf_{n}^{2}\mathring{m}^{2}(\theta)-\mathring{m}^{4}(\theta)\left\{\frac{(n^{2}-1)}{4(4\pi)^{2}}\left[\frac{1}{2}+\log\frac{\Lambda^{2}}{\mathring{m}^{2}(\theta)} \right ]\right.\\
&\left.+4n\left(nL^{r}_{6}-nL^{r}_{7}\tan^{2}\tfrac{\theta}{n}+L^{r}_{8}\right)\right\}+4B_{n}^{2}nm_{q}^{2}(2L^{r}_{8}-H^{r}_{2})+n_{Q}\frac{(eH)^{2}}{(4\pi)^{2}}\mathfrak{I}_{H}(\tfrac{\mathring{m}^{2}(\theta)}{eH})\\
\end{split}
\end{equation}
which is in agreement with Ref.~\cite{Guo:2015oxa} with the exception of $\theta$-independent contact terms, which we do not omit.
Using the definition of the topological susceptibility in Eq.~(\ref{eq:chitc40}), we get
\begin{equation}
\begin{split}
\label{eq:chit-nf}
\chi_{t}&=\frac{B_{n}m_{q}f_{n}^{2}}{n}+32B_{n}^{2}m_{q}^{2}(L^{r}_{6}+L^{r}_{7}+\tfrac{1}{n}L^{r}_{8})+\frac{2B_{n}^{2}m_{q}^{2}}{(4\pi)^{2}}\frac{n^{2}-1}{n^{2}}\log\frac{\Lambda^{2}}{\mathring{m}^{2}}\\
&-\frac{n_{Q}}{(4\pi)^{2}}\frac{2B_{n}m_{q}}{n^{2}}(eH) \mathcal{I}_{H,2}(\tfrac{\mathring{m}^{2}}{eH})\ ,
\end{split}
\end{equation}
and using the definition of the fourth cumulant in Eq.~(\ref{eq:chitc40}), we get
\begin{equation}
\begin{split}
\label{eq:c4-nf}
c_{4}=&-\frac{B_{n}m_{q}f_{n}^{2}}{n^{3}}-128B_{n}^{2}m_{q}^{2}\left[\tfrac{1}{n^{2}}(L^{r}_{6}+L^{r}_{7})+\tfrac{1}{n^{3}}L^{r}_{8}\right]+\frac{2B_{n}^{2}m_{q}^{2}}{(4\pi)^{2}}\frac{n^{2}-1}{n^{4}}\left(3-4\log\frac{\Lambda^{2}}{\mathring{m}^{2}}\right)\\
&+\frac{n_{Q}}{(4\pi)^{2}}\frac{2B_{n}m_{q}}{n^{4}}\left[eH\mathcal{I}_{H,2}(\tfrac{\mathring{m}^{2}}{eH})-6B_{n}m_{q}\mathcal{I}_{H,1}(\tfrac{\mathring{m}^{2}}{eH})\right]\ ,
\end{split}
\end{equation}
where the meson mass, $\mathring{m}\equiv\mathring{m}(0)$ and the integrals $\mathcal{I}_{H,2}$ and $\mathcal{I}_{H,1}$ are defined in Eqs.~$(\ref{eq:IH2})$ and $(\ref{eq:IH1})$ respectively. The $n$-flavor topological susceptibility and fourth cumulant with degenerate quarks is consistent with Eqs.~(\ref{eq:chit}) and (\ref{eq:c4}) after setting $n=3$ and $n_{Q}=2$, and making the replacements $B_{n}\rightarrow B_{0}$ and $f_{n}\rightarrow f_{0}$. Noting that the quark condensate is
\begin{align}
\langle\bar{q}q\rangle_{H}\equiv\sum_{f}\langle \bar{q}_{f}q_{f}\rangle_{H}=\frac{2B_{n}n_{Q}}{(4\pi)^{2}}(eH)\mathcal{I}_{H,2}(\tfrac{\mathring{m}^{2}}{eH})\ ,
\end{align}
we find a sum rule relating the shift of the topological susceptibility, $\chi_{t,H}$, in Eq.~(\ref{eq:chit-nf}) with the shift in the quark condensate,
\begin{align}
\chi_{t,H}=-m_{q}\frac{\langle\bar{q}q\rangle_{H}}{n^{2}}\ .
\end{align}
The sum rule is consistent with that of Eq.~(\ref{eq:sumrule-chitH}) for $n=3$ and degenerate quark masses. 
\section{Discussion}
\label{sec:discussion}
\noindent
In this section we use the analytical results of the previous section and plot the topological susceptibility and fourth cumulant in three-flavor $\chi$PT using physical masses for the pions and kaons. We will compare these results to that from two-flavor $\chi$PT calculated in Ref.~\cite{Adhikari:2021xra}. For the purposes of the comparison, we need to specify pion and kaon masses, decay constants and the quark masses, which we extract from the the Particle Data Group (PDG)~\cite{Zyla:2020zbs}, 
\begin{align}
\label{eq:quarkmasses}
m_{u}&=2.32\ {\rm MeV},\ m_{d}=3.71\ {\rm MeV},\ m_{s}=93.0\ {\rm MeV},\\
m_{\pi^{\pm}}&=139.6\ {\rm MeV},\ \sqrt{2}f_{\pi^{\pm}}=130.2\ {\rm MeV},\\ 
m_{K^{\pm}}&=493.7\ {\rm MeV},\ \sqrt{2}f_{K^{\pm}}=155.7\ {\rm MeV}\ .
\end{align}
We define the relative shifts of the topological susceptibility and the fourth cumulant using the following definition for the relative shift of a quantity $\mathcal{Q}$ as $\mathcal{R}_{\mathcal{Q}}=\frac{\mathcal{Q}_{H}}{\mathcal{Q}_{0}}$, where $\mathcal{Q}_{H}$ is the shift due to the magnetic field and at $\mathcal{O}(p^{4})$ we can choose the vacuum value, $\mathcal{Q}_{0}$ to be the tree level value. The NLO corrections of $\mathcal{Q}_{0}$ only contribute at $\mathcal{O}(p^{6})$. Using Eq.~(\ref{eq:chit}), in particular the first term, which is the tree-level topological susceptibility and the last term, which the $H$-dependent shift, we get for the relative shift of the topological susceptibility
\begin{equation}
\begin{split}
\label{eq:Rc4}
\mathcal{R}_{\chi_{t}}&=-\frac{\bar{m}(eH)}{(4\pi f)^{2}}\left[\left(\frac{1}{m_{u}}+\frac{1}{m_{d}}\right) \mathcal{I}_{H,2}(\tfrac{\mathring{m}_{\pi^{\pm}}^{2}}{eH})+\left(\frac{1}{m_{u}}+\frac{1}{m_{s}}\right)\mathcal{I}_{H,2}(\tfrac{\mathring{m}_{K^{\pm}}^{2}}{eH})\right]\ ,
\end{split}
\end{equation}
where $\mathring{m}_{\pi^{\pm}}$ and $\mathring{m}_{K^{\pm}}$ are the bare pion and kaon masses. Similarly, using Eq.~(\ref{eq:c4}), where the first term is the tree level susceptibility and the last two terms are the absolute shifts due to the magnetic field, we get
\begin{equation}
\begin{split}
\label{eq:Rc4} 
\mathcal{R}_{c_{4}}=&-\frac{m^{[3]}(eH)}{(4\pi f)^{2}}\left[\left(\frac{1}{m_{u}^{3}}+\frac{1}{m_{d}^{3}}\right)\mathcal{I}_{H,2}(\tfrac{\mathring{m}_{\pi^{\pm}}^{2}}{eH})+\left(\frac{1}{m_{u}^{3}}+\frac{1}{m_{s}^{3}}\right)\mathcal{I}_{H,2}(\tfrac{\mathring{m}_{K^{\pm}}^{2}}{eH})\right]\\
&+\frac{3B_{0}\bar{m}m^{[3]}}{(4\pi f)^{2}}\left[\frac{1}{\bar{m}_{ud}}\left(\frac{1}{m_{u}}+\frac{1}{m_{d}}\right)^{2}\mathcal{I}_{H,1}(\tfrac{\mathring{m}_{\pi^{\pm}}^{2}}{eH})+\frac{1}{\bar{m}_{us}}\left(\frac{1}{m_{u}}+\frac{1}{m_{s}}\right)^{2}\mathcal{I}_{H,1}(\tfrac{\mathring{m}_{K^{\pm}}^{2}}{eH})\right]\ ,
\end{split}
\end{equation}
where $\bar{m}$ is the three-flavor reduced mass and $\bar{m}_{ud}$ and $\bar{m}_{us}$ are defined in Eqs.~(\ref{eq:mbud}) and (\ref{eq:mbus}). In the limit of large strange quark mass, the three-flavor reduced mass $\bar{m}$ and $\bar{m}_{ud}$ both become two-flavor reduced mass, $(\frac{1}{m_{u}}+\frac{1}{m_{d}})^{-1}$, the mass $m^{[3]}\rightarrow(\frac{1}{m_{u}^{3}}+\frac{1}{m_{d}^{3}})^{-1}$ and the integrals $\mathcal{I}_{H,2}$ and $\mathcal{I}_{H,1}$ involving the bare kaon mass are exponentially suppressed. Finally identifying $B_{0}$ with $B$, which is correct at leading order~\cite{Gasser:1984gg} and the sub-leading corrections are next-to-next-to-leading order, we recover the two-flavor relative shifts from \cite{Adhikari:2021xra} from the three-flavor results.
\begin{figure}
	\centering
	\begin{subfigure}[b]{0.48\textwidth}
	\includegraphics[width=\textwidth]{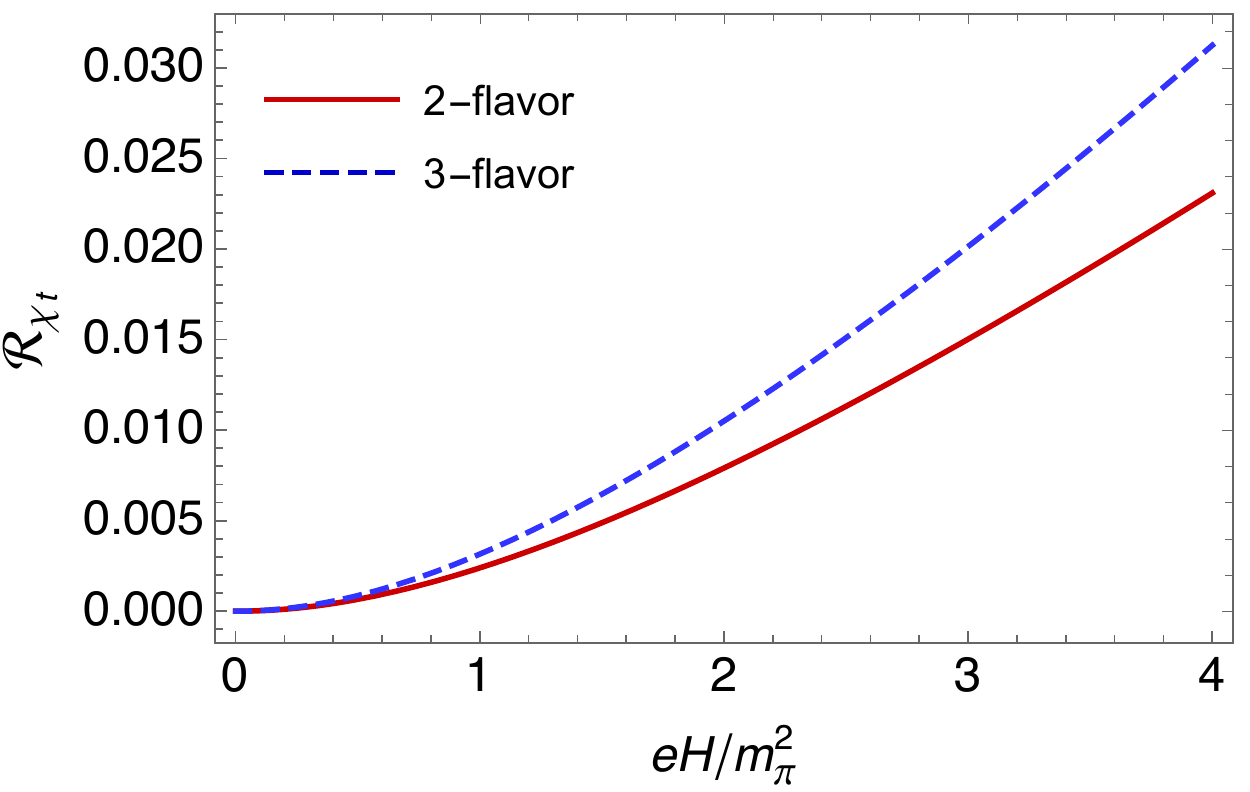}
	\end{subfigure}
	\begin{subfigure}[b]{0.48\textwidth}
	\includegraphics[width=\textwidth]{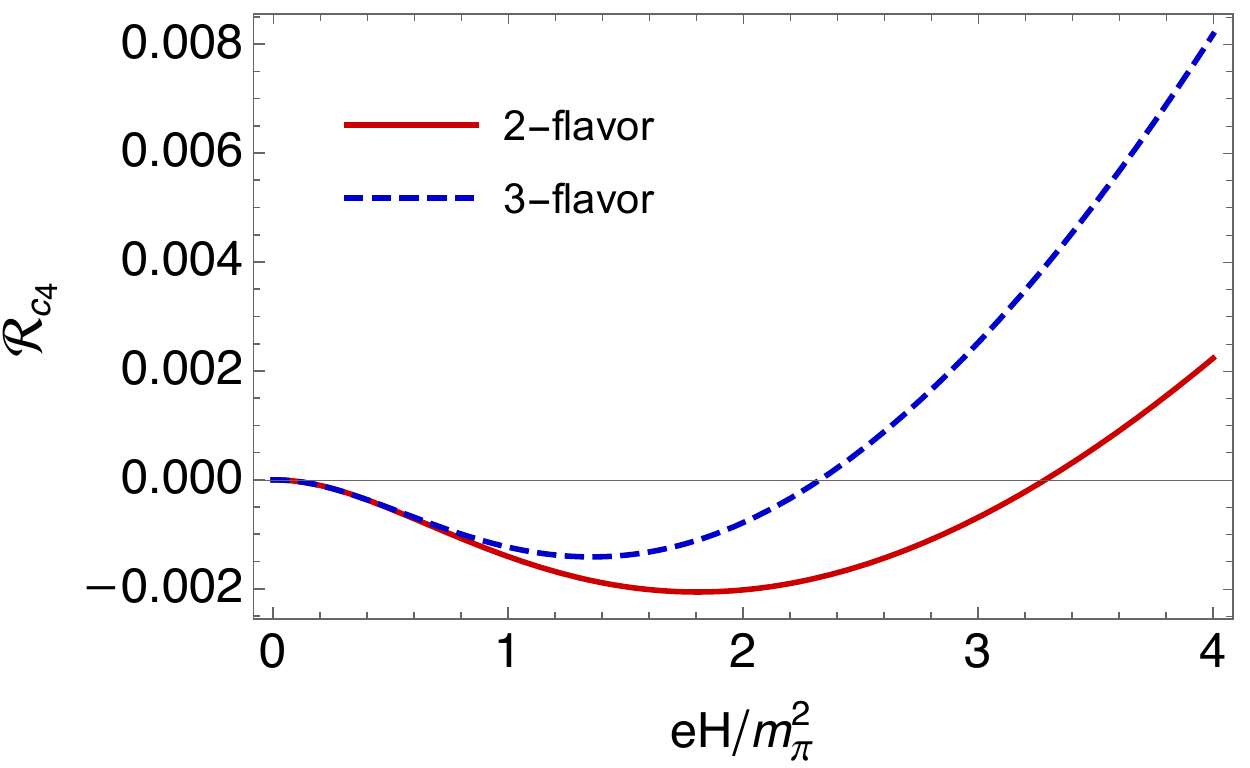}
	\end{subfigure}
\caption{Plots of the relative shift in the topological susceptibility (fourth cumulant) relative to their $H=0$ vacuum values as a function the background magnetic field on the left (right) panel in two-and-three flavor $\chi$PT.}
	\label{fig:rchitc4}
\end{figure}

In Fig.~\ref{fig:rchitc4}, we plot the relative shifts of the topological susceptibility and fourth cumulants in two-and-three-flavor on the left and right panels respectively. The solid, red line shows the results in two-flavor $\chi$PT and the dashed, blue line indicates the results in three-flavor $\chi$PT. In order to generate the plots, we have calculated the bare pion mass and pion decay constant using the renormalized pion mass, kaon mass and the pion decay constant, the expression for which can be found in \ref{sec:renormalized}. We use the central values of the following three-flavor LECs
\begin{align}
10^{3}{L}_{1}^r&=1.0\pm 0.1\ ,&
10^{3}{L}_{2}^r&=1.6\pm 0.2\ ,& 10^{3}{L}_{3}^r&=-3.8\pm 0.3\ ,&
10^{3}{L}_{4}^r&=0.0 \pm 0.3\ ,\\
10^{3}{L}_{5}^r&=1.2 \pm 0.1\ ,&10^{3}{L}_{6}^r&=0.0 \pm 0.4\ ,& 10^{3}{L}_{7}^r&=-0.4 \pm 0.2\ , & 10^{3}{L}_{8}^r&=0.5 \pm 0.2\ ,
\label{eq:LEC3f}
\end{align}
defined at the scale $\Lambda=\sqrt{4\pi e^{-\gamma_E}}m_{\rho}$, where $m_{\rho}=0.77\ {\rm GeV}$ and the central values of the following two-flavor LECs,
\begin{align}
\label{eq:LEC2f}
\bar{l}_{1}&=-0.4\pm0.6\ ,&\bar{l}_{2}&=4.3\pm0.1\ ,&\bar{l}_{3}&=2.9\pm2.4\ ,&\bar{l}_{4}&=4.4\pm0.2\ ,
\end{align}
to calculate the bare mass and decay constants. Using Eqs.~(\ref{eq:pionmass3f}),(\ref{eq:kaonmass3f}) and (\ref{eq:piondecay3f}), we get in the three-flavor case
\begin{align}
\mathring{m}_{\pi}&=140.1\ {\rm MeV}\ ,\ \mathring{m}_{K}=533.0\ {\rm MeV}\ ,\ f=76.5\ {\rm MeV}\ \ ,
\end{align}
and using  Eqs.~(\ref{eq:pionmass2f}) and (\ref{eq:piondecay2f}) in the two-flavor case
\begin{align}
\mathring{m}_{\pi}&=141.3\ {\rm MeV}\ ,\  f=85.6\ {\rm MeV}\ .
\end{align}
We find that the enhancement of the topological susceptibility in Fig.~\ref{fig:rchitc4} is greater in three-flavor $\chi$PT for all values of the magnetic field. For instance at $eH=4m_{\pi}^{2}$, the relative shift of the topological susceptibility is approximately $0.03$ in the three-flavor case while it is approximately $0.02$ in the two-flavor case. The relative shift of the fourth cumulant, on the other hand, is comparable at low magnetic field up to $eH\approx0.8m_{\pi}^{2}$ with the three-flavor result less suppressed for larger fields. For values of the magnetic field larger than $eH\approx0.8m_{\pi}^{2}$ the suppression of two-flavor fourth cumulant is greater. The three-flavor fourth cumulant is enhanced beginning at a magnetic field, $eH\approx2.2m_{\pi}^{2}$, with the two-flavor fourth cumulant being enhanced starting at a larger magnetic field, $eH\approx3.3m_{\pi}^{2}$. In the region where both fourth cumulants are enhanced, the relative shift in the three-flavor case is significantly larger than the two-flavor case. The fourth cumulant is suppressed at low magnetic fields due to the structure of the relative shift in Eq.~(\ref{eq:Rc4}) which contains two negative definite integrals $\mathcal{I}_{n,1}$ and $\mathcal{I}_{n,2}$ defined in Eq.~(\ref{eq:IHn}). At low magnetic fields, the integrals are exponentially suppressed and the negative contribution proportional to $\mathcal{I}_{H,1}$ dominates while at high magnetic fields the integral $\mathcal{I}_{H,2}$ dominates. Since the terms proportional to $\mathcal{I}_{n,2}$ come with negative coefficients, the relative shift of the fourth cumulant becomes positive.
\begin{figure}
	\centering
	\begin{subfigure}[b]{0.45\textwidth}
	\includegraphics[width=\textwidth]{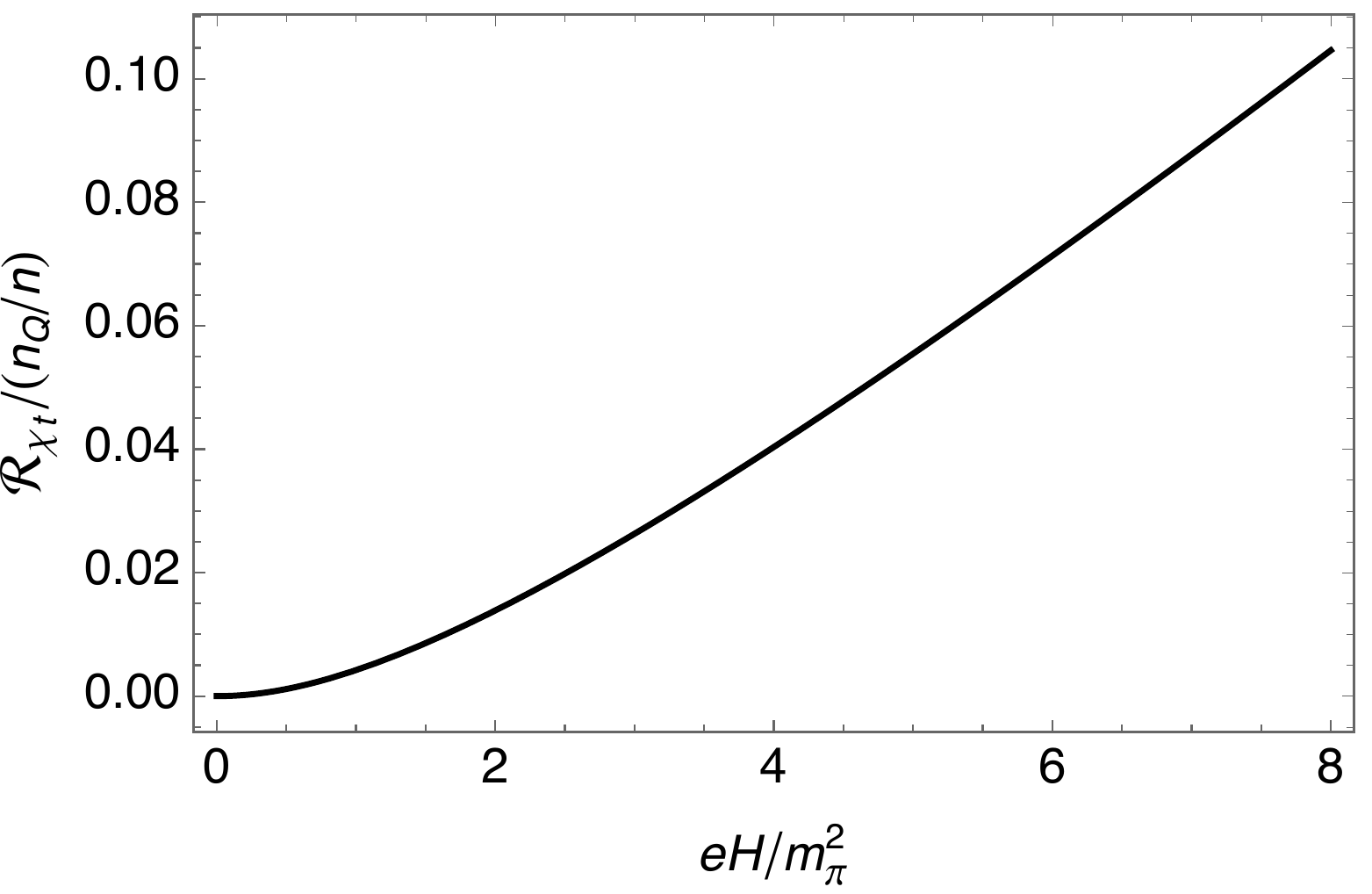}
	\end{subfigure}
	\begin{subfigure}[b]{0.45\textwidth}
	\includegraphics[width=\textwidth]{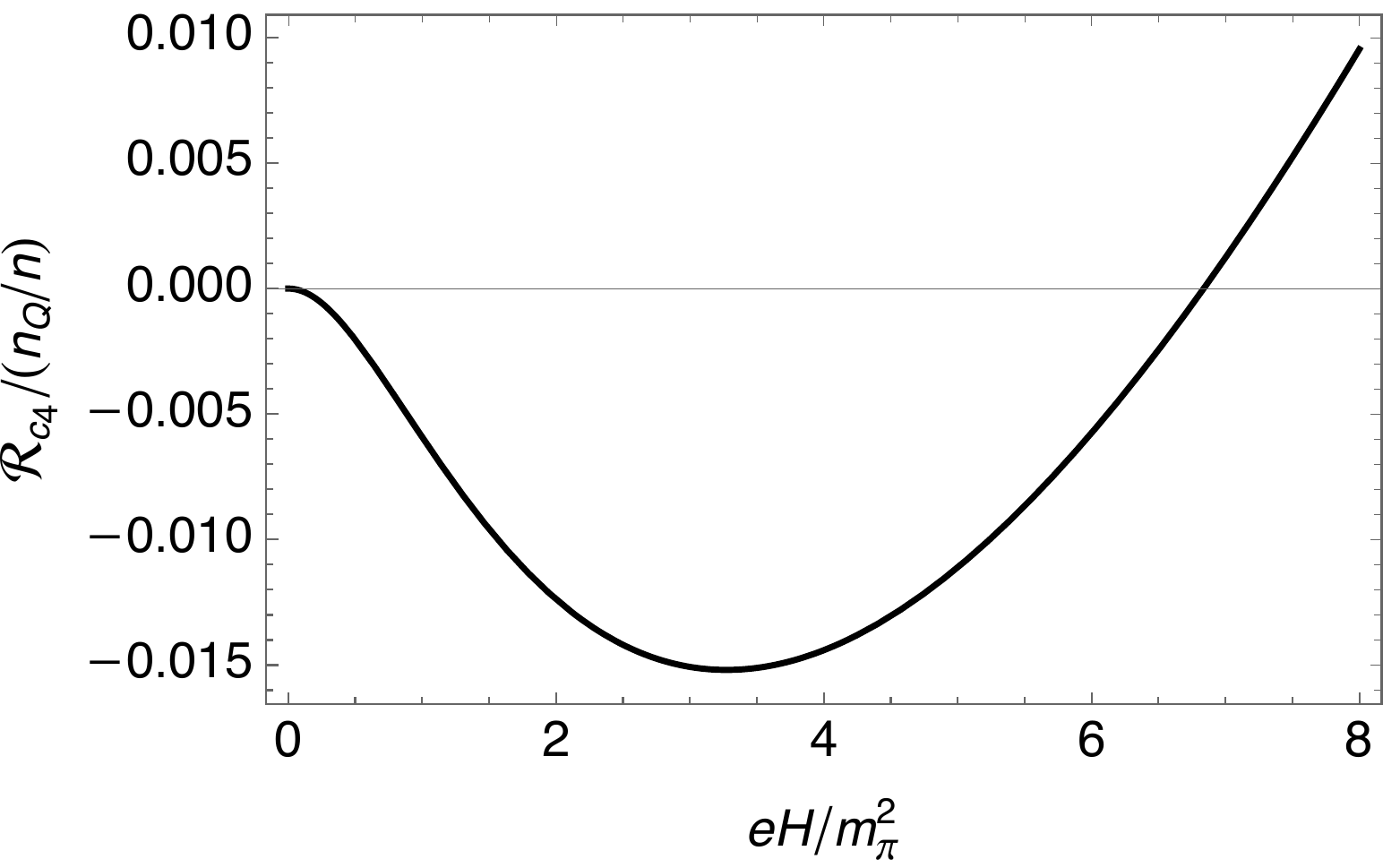}
	\end{subfigure}
\caption{Plots of the relative shift in the topological susceptibility (fourth cumulant) relative to their $H=0$ vacuum values (normalized by $n_{Q}/n$) as a function the background magnetic field on the left (right) panel in $n$-flavor $\chi$PT.}
	\label{fig:rchitc4-nf}
\end{figure}

In Fig.~\ref{fig:rchitc4-nf}, we plot the relative shifts of the topological susceptibility and fourth cumulants normalized by the ratio $n_{Q}/n$, where $n_{Q}$ is the number of charged pairs and $n$ is the number of flavors. The origin of this ratio can be understood by considering the tree level and the $H$-dependent next-to-leading order contributions, which are the only quantities that appear in the relative shift. At tree-level the topological susceptibility scales with the number of flavors as $1/n$ while the $H$-dependent shift scales as $1/n^{2}$ in addition to scaling linearly with $n_{Q}$. As such, the relative shift of the topological susceptibility scales as $n_{Q}/n$. In the two-flavor scenario with degenerate quarks the ratio is $\tfrac{1}{2}$ while in the three-flavor case with degenerate quarks the ratio is $\tfrac{2}{3}$ suggesting that the susceptibility is a $\tfrac{1}{3}$rd larger in the three-flavor case. This is not even approximately borne out in Fig.~\ref{fig:rchitc4} since the strange quark masses are considerably larger than the masses of the up and the down quarks.

We can use Eqs.~(\ref{eq:chit-nf}) and (\ref{eq:c4-nf}), to calculate the relative shifts of the cumulants,
\begin{align}
\mathcal{R}_{\chi_{t},n}&=-\frac{2n_{Q}(eH)}{n(4\pi f)^{2}}\mathcal{I}_{H,2}(\tfrac{\mathring{m}^{2}}{eH})\ ,\\
\mathcal{R}_{c_{4},n}&=-\frac{2n_{Q}}{n(4\pi f)^{2}}\left[eH\mathcal{I}_{H,2}(\tfrac{\mathring{m}^{2}}{eH})-6B_{n}m_{q}\mathcal{I}_{H,1}(\tfrac{\mathring{m}^{2}}{eH})\right]\ ,
\end{align}
where $n$ is the number of flavors and $n_{Q}$ is the number of charged pairs of mesons, which depends on the charge matrix $Q$. 
Unsurprisingly the relative shifts scale linearly with $n_{Q}$. As such in our plots in Fig.~\ref{fig:rchitc4-nf}, we normalize the cumulants by $n_{Q}$. Since the expression is only valid for degenerate quarks, we use $m_{q}=\tfrac{m_{u}+m_{d}}{2}=3.515\ {\rm MeV}$ for the quark mass consistent with Eq.~(\ref{eq:quarkmasses}). We use $\sqrt{2}f_{n}=130.2\ {\rm MeV}$, $\mathring{m}=139.6\ {\rm MeV}$ and $B_{n}=2772.14\ {\rm MeV}$, where the latter is found using the Gell-Mann-Oakes-Renner relation. The relative shift of the topological susceptibility normalized by $n_{Q}/n$ is shown in the left panel of Fig.~\ref{fig:rchitc4-nf} and the shift is comparable to results from two-flavor and three-flavor $\chi$PT with non-degenerate quarks in Fig.~\ref{fig:rchitc4} modulo the factor of $n_{Q}/n$. However, the relative shift of the fourth cumulant is markedly different. While the fourth cumulant is suppressed at low magnetic fields, it is not enhanced until a critical magnetic field of $eH_{c}\approx6.85m_{\pi}^{2}$, which is considerably higher than the results in two and three-flavor $\chi$PT with degenerate quarks where the critical magnetic field is less than half as large. This suggests that the critical magnetic field is quite sensitive to the difference in quark masses.
\section{Conclusion}
\label{sec:conclusion}
In this paper, we have studied the shift in the first two topological cumulants in the presence of a background magnetic field in three-flavor $\chi$PT with non-degenerate quark masses and $n$-flavor $\chi$PT with degenerate quark masses. We have compared the relative shifts of the cumulants to that from two-flavor $\chi$PT finding that the shift in the topological susceptibility is enhanced in the three-flavor case compared to two-flavors. The fourth cumulant, however, was suppressed comparably at low fields with the large magnetic field enhancement occurring at a higher magnetic field in the two-flavor case and the enhancement being lower in the two-flavor case. We also studied the topological cumulants for degenerate $n$-flavor $\chi$PT finding that the topological susceptibility is enhanced for magnetic fields comparably to two and three-flavor $\chi$PT with the fourth cumulant being significantly different. Our results are model-independent and explore the QCD vacuum with $\theta=0$, where there is no sign problem unlike at finite $\theta$. Consequently, our results can be compared to future lattice calculations.

\section*{Acknowledgements} 
\noindent 
P.A. would like to acknowledge the support of St. Olaf College start up funds and CCNY for their hospitality during the latter stages of this work. P.A. also acknowledges helpful discussions and collaboration with J.O. Andersen and Inga Str\"{u}mke on magnetic catalysis and magnetization within three-flavor $\chi$PT.

\appendix
\section{Useful Integrals}
\label{sec:integrals}
\noindent
Here we list all the relevant Schwinger integrals relevant to a finite $H$ calculation beginning with the one-loop effective potential of a pair of charged bosons with mass $m$ and charge $\pm e$,
\begin{equation}
\begin{split}
\label{eq:Schwinger-integral}
I_{H}^{\rm fin}(m)&=\frac{(eH)^{2}}{(4\pi)^{2}}\mathfrak{I}_{H}(\tfrac{m^{2}}{eH})\\
\mathfrak{I}_{H}(x)&=-\int_{0}^{\infty}dz\ \frac{e^{-xz}}{z^{3}}\left[\frac{z}{\sinh z}-1+\frac{z^{2}}{6}\right]\\
&=4\zeta^{(1,0)}(-1,\tfrac{x+1}{2})+(\tfrac{x}{2})^{2}(1-2\log \tfrac{x}{2})+\tfrac{1}{6}(\log \tfrac{x}{2}+1)\ ,
\end{split}
\end{equation}
 where $x=\frac{m^{2}}{eH}$ and $\zeta(s,a)$ is the Hurwitz zeta function with the numbers in the subscripts indicating the number of derivatives with respect to $s$ and $a$ in that order. 
Additionally, we require further Schwinger integrals to characterize the quark condensates, susceptibility and the topological cumulants,
\begin{equation}
\begin{split}
\label{eq:IHn}
\mathcal{I}_{H,n}(x)&=\int_{0}^{\infty}dz\frac{e^{-xz}}{z^{n}}\left(\frac{z}{\sinh z}-1\right)\ .
\end{split}
\end{equation}
For $n=2,1,0,$ and $-1$, they are 
\begin{align}
\label{eq:IH2}
\mathcal{I}_{H,2}(x)&=2\zeta^{(1,0)}(0,\tfrac{x+1}{2})-x(\log \tfrac{x}{2}-1)\\
\label{eq:IH1}
\mathcal{I}_{H,1}(x)&=\log\tfrac{x}{2}-\psi_{0}\left (\tfrac{x+1}{2} \right)\\
\label{eq:IH0}
\mathcal{I}_{H,0}(x)&=-\tfrac{1}{x}+\tfrac{1}{2}\psi_{1}(\tfrac{x+1}{2})\\
\label{eq:IH-1}
\mathcal{I}_{H,-1}(x)&=-\tfrac{1}{x^{2}}-\tfrac{1}{4}\psi_{2}(\tfrac{x+1}{2})\ ,
\end{align}
where $\psi_{n}(x)$ is the polygamma function, which is defined in terms of the derivatives of the $\Gamma(x)$ function as
\begin{align}
\psi_{n}(x)=\frac{d^{n+1}}{dx^{n+1}}\log\Gamma(x)\ .
\end{align}
$\mathcal{I}_{H,n}(x)$ is negative definite and vanishes in the absence of the external magnetic field.
\section{1-loop renormalized masses and decay constants}
\label{sec:renormalized}
\noindent
Since the mixing angle, $\epsilon$ is small $\approx1.16\times10^{-2}$, we expect that the bare quantities can calculated with reasonable accuracy using the three-flavor pion and kaon masses and the decay constants in the isospin limit. The three-flavor pion and kaon masses are~\cite{Gasser:1984gg}
\begin{align}
\nonumber
\label{eq:pionmass3f}
m_{\pi}^2&=\mathring{m}_{\pi}^2\left[1-\left(8{L}_4^r+8{L}_5^r-16{L}_6^r-16{L}_8^r+{1\over2(4\pi)^2}\log{\Lambda^2\over \mathring{m}_{\pi}^2}\right){\mathring{m}_{\pi}^2\over f^2}-({L}_4^r-2{L}_6^r){16\mathring{m}_{K}^2\over f^2}\right.\\
&\left.+{\mathring{m}_{\eta}^2\over6(4\pi)^2f^2}\log{\Lambda^2\over \mathring{m}_{\eta}^2}\right]\ ,\\
\label{eq:kaonmass3f}
m_{K}^2&=\mathring{m}_{K}^2\left[1-\left({L}_4^r-2{L}_6^r\right){8\mathring{m}_{\pi}^2\over f^2}-(2L_4^r+{L}_5^r-4L_6^r-2{L}_8^r){8\mathring{m}_{\eta}^2\over f^2}-{\mathring{m}_{\eta}^2\over3(4\pi)^2f^2}\log{\Lambda^2\over \mathring{m}_{\eta}^2}\right]\ ,
\end{align}
and the three-flavor pion and kaon decay constants, $f_{\pi}$ and $f_{K}$ are respectively~\cite{Gasser:1984gg}
\begin{align}
\label{eq:piondecay3f}
f_{\pi}^2&=f^2\left[1+\left(8{L}_4^r+8L_5^r+{2\over(4\pi)^2}\log{\Lambda^2\over \mathring{m}_{\pi}^2}
\right){\mathring{m}_{\pi}^2\over f^2}+\left(16L_4^r+{1\over(4\pi)^2}\log{\Lambda^2\over \mathring{m}_{K}^2}\right){\mathring{m}_{K}^2\over f^2}\right]\\
\nonumber
f_K^2&=f^2\left[1+\left(12L_4^r+{3\over4(4\pi)^2}\log{\Lambda^2\over \mathring{m}_{\pi}^2}\right){\mathring{m}_{\pi}^2\over f^2}+\left(8L_5^r+{3\over2(4\pi)^2}\log{\Lambda^2\over \mathring{m}_{K}^2}\right){\mathring{m}_{K}^2\over f^2}+\right.\\
&\left.\left(12L_4^r+{3\over4(4\pi)^2}\log{\Lambda^2\over \mathring{m}_{\eta}^2}\right){\mathring{m}_{\eta}^2\over f^2}\right]\ ,
\end{align}
where $L^{r}_{i}$ are the low energy constants defined at the $\overline{MS}$ scale $\Lambda$. In the two-flavor case, the pion mass and the decay constants are
\begin{align}
\label{eq:pionmass2f}
m_{\pi}^{2}&=\mathring{m}_{\pi}^2\left[1-{\mathring{m}_{\pi}^2\over2(4\pi)^2f^2}\bar{l}_3\right]\\
\label{eq:piondecay2f}
f_{\pi}^2&=f^2\left[1+{2\mathring{m}_{\pi}^2\over(4\pi)^2f^2}\bar{l}_4\right]\ ,
\end{align}
where $\bar{l}_{i}$ are the rescaled low energy constants defined at the scale of the bare pion mass~\cite{Gasser:1983yg}.
\bibliographystyle{apsrev4-1}
\bibliography{/Users/prabal7e7/Documents/Research/bib}

\begin{thebibliography}{22}%
\makeatletter
\providecommand \@ifxundefined [1]{%
 \@ifx{#1\undefined}
}%
\providecommand \@ifnum [1]{%
 \ifnum #1\expandafter \@firstoftwo
 \else \expandafter \@secondoftwo
 \fi
}%
\providecommand \@ifx [1]{%
 \ifx #1\expandafter \@firstoftwo
 \else \expandafter \@secondoftwo
 \fi
}%
\providecommand \natexlab [1]{#1}%
\providecommand \enquote  [1]{``#1''}%
\providecommand \bibnamefont  [1]{#1}%
\providecommand \bibfnamefont [1]{#1}%
\providecommand \citenamefont [1]{#1}%
\providecommand \href@noop [0]{\@secondoftwo}%
\providecommand \href [0]{\begingroup \@sanitize@url \@href}%
\providecommand \@href[1]{\@@startlink{#1}\@@href}%
\providecommand \@@href[1]{\endgroup#1\@@endlink}%
\providecommand \@sanitize@url [0]{\catcode `\\12\catcode `\$12\catcode
  `\&12\catcode `\#12\catcode `\^12\catcode `\_12\catcode `\%12\relax}%
\providecommand \@@startlink[1]{}%
\providecommand \@@endlink[0]{}%
\providecommand \url  [0]{\begingroup\@sanitize@url \@url }%
\providecommand \@url [1]{\endgroup\@href {#1}{\urlprefix }}%
\providecommand \urlprefix  [0]{URL }%
\providecommand \Eprint [0]{\href }%
\providecommand \doibase [0]{http://dx.doi.org/}%
\providecommand \selectlanguage [0]{\@gobble}%
\providecommand \bibinfo  [0]{\@secondoftwo}%
\providecommand \bibfield  [0]{\@secondoftwo}%
\providecommand \translation [1]{[#1]}%
\providecommand \BibitemOpen [0]{}%
\providecommand \bibitemStop [0]{}%
\providecommand \bibitemNoStop [0]{.\EOS\space}%
\providecommand \EOS [0]{\spacefactor3000\relax}%
\providecommand \BibitemShut  [1]{\csname bibitem#1\endcsname}%
\let\auto@bib@innerbib\@empty
\bibitem [{\citenamefont {Crewther}(1977)}]{Crewther:1977ce}%
  \BibitemOpen
  \bibfield  {author} {\bibinfo {author} {\bibfnamefont {R.~J.}\ \bibnamefont
  {Crewther}},\ }\href {\doibase 10.1016/0370-2693(77)90675-X} {\bibfield
  {journal} {\bibinfo  {journal} {Phys. Lett. B}\ }\textbf {\bibinfo {volume}
  {70}},\ \bibinfo {pages} {349} (\bibinfo {year} {1977})}\BibitemShut
  {NoStop}%
\bibitem [{\citenamefont {Witten}(1979)}]{Witten:1979vv}%
  \BibitemOpen
  \bibfield  {author} {\bibinfo {author} {\bibfnamefont {E.}~\bibnamefont
  {Witten}},\ }\href {\doibase 10.1016/0550-3213(79)90031-2} {\bibfield
  {journal} {\bibinfo  {journal} {Nucl. Phys. B}\ }\textbf {\bibinfo {volume}
  {156}},\ \bibinfo {pages} {269} (\bibinfo {year} {1979})}\BibitemShut
  {NoStop}%
\bibitem [{\citenamefont {Veneziano}(1979)}]{Veneziano:1979ec}%
  \BibitemOpen
  \bibfield  {author} {\bibinfo {author} {\bibfnamefont {G.}~\bibnamefont
  {Veneziano}},\ }\href {\doibase 10.1016/0550-3213(79)90332-8} {\bibfield
  {journal} {\bibinfo  {journal} {Nucl. Phys. B}\ }\textbf {\bibinfo {volume}
  {159}},\ \bibinfo {pages} {213} (\bibinfo {year} {1979})}\BibitemShut
  {NoStop}%
\bibitem [{\citenamefont {Di~Vecchia}\ and\ \citenamefont
  {Veneziano}(1980)}]{DiVecchia:1980yfw}%
  \BibitemOpen
  \bibfield  {author} {\bibinfo {author} {\bibfnamefont {P.}~\bibnamefont
  {Di~Vecchia}}\ and\ \bibinfo {author} {\bibfnamefont {G.}~\bibnamefont
  {Veneziano}},\ }\href {\doibase 10.1016/0550-3213(80)90370-3} {\bibfield
  {journal} {\bibinfo  {journal} {Nucl. Phys. B}\ }\textbf {\bibinfo {volume}
  {171}},\ \bibinfo {pages} {253} (\bibinfo {year} {1980})}\BibitemShut
  {NoStop}%
\bibitem [{\citenamefont {'t~Hooft}(1976{\natexlab{a}})}]{tHooft:1976rip}%
  \BibitemOpen
  \bibfield  {author} {\bibinfo {author} {\bibfnamefont {G.}~\bibnamefont
  {'t~Hooft}},\ }\href {\doibase 10.1103/PhysRevLett.37.8} {\bibfield
  {journal} {\bibinfo  {journal} {Phys. Rev. Lett.}\ }\textbf {\bibinfo
  {volume} {37}},\ \bibinfo {pages} {8} (\bibinfo {year}
  {1976}{\natexlab{a}})}\BibitemShut {NoStop}%
\bibitem [{\citenamefont {'t~Hooft}(1976{\natexlab{b}})}]{tHooft:1976snw}%
  \BibitemOpen
  \bibfield  {author} {\bibinfo {author} {\bibfnamefont {G.}~\bibnamefont
  {'t~Hooft}},\ }\href {\doibase 10.1103/PhysRevD.14.3432} {\bibfield
  {journal} {\bibinfo  {journal} {Phys. Rev. D}\ }\textbf {\bibinfo {volume}
  {14}},\ \bibinfo {pages} {3432} (\bibinfo {year} {1976}{\natexlab{b}})},\
  \bibinfo {note} {[Erratum: Phys.Rev.D 18, 2199 (1978)]}\BibitemShut {NoStop}%
\bibitem [{\citenamefont {Weinberg}(1975)}]{Weinberg:1975ui}%
  \BibitemOpen
  \bibfield  {author} {\bibinfo {author} {\bibfnamefont {S.}~\bibnamefont
  {Weinberg}},\ }\href {\doibase 10.1103/PhysRevD.11.3583} {\bibfield
  {journal} {\bibinfo  {journal} {Phys. Rev. D}\ }\textbf {\bibinfo {volume}
  {11}},\ \bibinfo {pages} {3583} (\bibinfo {year} {1975})}\BibitemShut
  {NoStop}%
\bibitem [{\citenamefont {Fujikawa}(1979)}]{Fujikawa:1979ay}%
  \BibitemOpen
  \bibfield  {author} {\bibinfo {author} {\bibfnamefont {K.}~\bibnamefont
  {Fujikawa}},\ }\href {\doibase 10.1103/PhysRevLett.42.1195} {\bibfield
  {journal} {\bibinfo  {journal} {Phys. Rev. Lett.}\ }\textbf {\bibinfo
  {volume} {42}},\ \bibinfo {pages} {1195} (\bibinfo {year}
  {1979})}\BibitemShut {NoStop}%
\bibitem [{\citenamefont {Srednicki}(2007)}]{srednicki2007quantum}%
  \BibitemOpen
  \bibfield  {author} {\bibinfo {author} {\bibfnamefont {M.}~\bibnamefont
  {Srednicki}},\ }\href@noop {} {\emph {\bibinfo {title} {Quantum field
  theory}}}\ (\bibinfo  {publisher} {Cambridge University Press},\ \bibinfo
  {year} {2007})\BibitemShut {NoStop}%
\bibitem [{\citenamefont {Adhikari}(2022{\natexlab{a}})}]{Adhikari:2021xra}%
  \BibitemOpen
  \bibfield  {author} {\bibinfo {author} {\bibfnamefont {P.}~\bibnamefont
  {Adhikari}},\ }\href@noop {} {\bibfield  {journal} {\bibinfo  {journal}
  {Nucl. Phys. B}\ }\textbf {\bibinfo {volume} {974}} (\bibinfo {year}
  {2022}{\natexlab{a}})},\ \Eprint {http://arxiv.org/abs/2111.06196}
  {arXiv:2111.06196} \BibitemShut {NoStop}%
\bibitem [{\citenamefont {Lu}\ \emph {et~al.}(2020)\citenamefont {Lu},
  \citenamefont {Du}, \citenamefont {Guo}, \citenamefont {Mei{\ss}ner},\ and\
  \citenamefont {Vonk}}]{Lu_2020}%
  \BibitemOpen
  \bibfield  {author} {\bibinfo {author} {\bibfnamefont {Z.-Y.}\ \bibnamefont
  {Lu}}, \bibinfo {author} {\bibfnamefont {M.-L.}\ \bibnamefont {Du}}, \bibinfo
  {author} {\bibfnamefont {F.-K.}\ \bibnamefont {Guo}}, \bibinfo {author}
  {\bibfnamefont {U.-G.}\ \bibnamefont {Mei{\ss}ner}}, \ and\ \bibinfo {author}
  {\bibfnamefont {T.}~\bibnamefont {Vonk}},\ }\href {\doibase
  10.1007/jhep05(2020)001} {\bibfield  {journal} {\bibinfo  {journal} {JHEP}\
  }\textbf {\bibinfo {volume} {2020}} (\bibinfo {year} {2020}),\
  10.1007/jhep05(2020)001}\BibitemShut {NoStop}%
\bibitem [{\citenamefont {Adhikari}(2022{\natexlab{b}})}]{Adhikari:2021lbl}%
  \BibitemOpen
  \bibfield  {author} {\bibinfo {author} {\bibfnamefont {P.}~\bibnamefont
  {Adhikari}},\ }\href@noop {} {\bibfield  {journal} {\bibinfo  {journal}
  {Phys. Lett. B}\ }\textbf {\bibinfo {volume} {825}} (\bibinfo {year}
  {2022}{\natexlab{b}})},\ \Eprint {http://arxiv.org/abs/2103.05048}
  {arXiv:2103.05048 [hep-ph]} \BibitemShut {NoStop}%
\bibitem [{\citenamefont {Vicari}\ and\ \citenamefont
  {Panagopoulos}(2009)}]{VICARI200993}%
  \BibitemOpen
  \bibfield  {author} {\bibinfo {author} {\bibfnamefont {E.}~\bibnamefont
  {Vicari}}\ and\ \bibinfo {author} {\bibfnamefont {H.}~\bibnamefont
  {Panagopoulos}},\ }\href {\doibase
  https://doi.org/10.1016/j.physrep.2008.10.001} {\bibfield  {journal}
  {\bibinfo  {journal} {Physics Reports}\ }\textbf {\bibinfo {volume} {470}},\
  \bibinfo {pages} {93 } (\bibinfo {year} {2009})}\BibitemShut {NoStop}%
\bibitem [{\citenamefont {Gasser}\ and\ \citenamefont
  {Leutwyler}(1985)}]{Gasser:1984gg}%
  \BibitemOpen
  \bibfield  {author} {\bibinfo {author} {\bibfnamefont {J.}~\bibnamefont
  {Gasser}}\ and\ \bibinfo {author} {\bibfnamefont {H.}~\bibnamefont
  {Leutwyler}},\ }\href {\doibase 10.1016/0550-3213(85)90492-4} {\bibfield
  {journal} {\bibinfo  {journal} {Nucl. Phys.}\ }\textbf {\bibinfo {volume}
  {B250}},\ \bibinfo {pages} {465} (\bibinfo {year} {1985})}\BibitemShut
  {NoStop}%
\bibitem [{\citenamefont {Scherer}\ and\ \citenamefont
  {Schindler}(2011)}]{scherer2011primer}%
  \BibitemOpen
  \bibfield  {author} {\bibinfo {author} {\bibfnamefont {S.}~\bibnamefont
  {Scherer}}\ and\ \bibinfo {author} {\bibfnamefont {M.~R.}\ \bibnamefont
  {Schindler}},\ }\href@noop {} {\emph {\bibinfo {title} {A primer for chiral
  perturbation theory}}},\ Vol.\ \bibinfo {volume} {830}\ (\bibinfo
  {publisher} {Springer},\ \bibinfo {year} {2011})\BibitemShut {NoStop}%
\bibitem [{\citenamefont {Mao}\ and\ \citenamefont {Chiu}(2009)}]{Mao:2009sy}%
  \BibitemOpen
  \bibfield  {author} {\bibinfo {author} {\bibfnamefont {Y.-Y.}\ \bibnamefont
  {Mao}}\ and\ \bibinfo {author} {\bibfnamefont {T.-W.}\ \bibnamefont {Chiu}}
  (\bibinfo {collaboration} {TWQCD}),\ }\href {\doibase
  10.1103/PhysRevD.80.034502} {\bibfield  {journal} {\bibinfo  {journal} {Phys.
  Rev. D}\ }\textbf {\bibinfo {volume} {80}},\ \bibinfo {pages} {034502}
  (\bibinfo {year} {2009})},\ \Eprint {http://arxiv.org/abs/0903.2146}
  {arXiv:0903.2146 [hep-lat]} \BibitemShut {NoStop}%
\bibitem [{\citenamefont {Guo}\ and\ \citenamefont
  {Mei\ss{}ner}(2015)}]{Guo:2015oxa}%
  \BibitemOpen
  \bibfield  {author} {\bibinfo {author} {\bibfnamefont {F.-K.}\ \bibnamefont
  {Guo}}\ and\ \bibinfo {author} {\bibfnamefont {U.-G.}\ \bibnamefont
  {Mei\ss{}ner}},\ }\href {\doibase 10.1016/j.physletb.2015.07.076} {\bibfield
  {journal} {\bibinfo  {journal} {Phys. Lett. B}\ }\textbf {\bibinfo {volume}
  {749}},\ \bibinfo {pages} {278} (\bibinfo {year} {2015})},\ \Eprint
  {http://arxiv.org/abs/1506.05487} {arXiv:1506.05487 [hep-ph]} \BibitemShut
  {NoStop}%
\bibitem [{\citenamefont {Schwinger}(1951)}]{Schwinger:1951nm}%
  \BibitemOpen
  \bibfield  {author} {\bibinfo {author} {\bibfnamefont {J.~S.}\ \bibnamefont
  {Schwinger}},\ }\href {\doibase 10.1103/PhysRev.82.664} {\bibfield  {journal}
  {\bibinfo  {journal} {Phys. Rev.}\ }\textbf {\bibinfo {volume} {82}},\
  \bibinfo {pages} {664} (\bibinfo {year} {1951})}\BibitemShut {NoStop}%
\bibitem [{\citenamefont {Bernard}\ \emph {et~al.}(2012)\citenamefont
  {Bernard}, \citenamefont {Descotes-Genon},\ and\ \citenamefont
  {Toucas}}]{Bernard:2012fw}%
  \BibitemOpen
  \bibfield  {author} {\bibinfo {author} {\bibfnamefont {V.}~\bibnamefont
  {Bernard}}, \bibinfo {author} {\bibfnamefont {S.}~\bibnamefont
  {Descotes-Genon}}, \ and\ \bibinfo {author} {\bibfnamefont {G.}~\bibnamefont
  {Toucas}},\ }\href {\doibase 10.1007/JHEP06(2012)051} {\bibfield  {journal}
  {\bibinfo  {journal} {JHEP}\ }\textbf {\bibinfo {volume} {06}},\ \bibinfo
  {pages} {051} (\bibinfo {year} {2012})},\ \Eprint
  {http://arxiv.org/abs/1203.0508} {arXiv:1203.0508 [hep-ph]} \BibitemShut
  {NoStop}%
\bibitem [{\citenamefont {G\'omez~Nicola}\ \emph {et~al.}(2019)\citenamefont
  {G\'omez~Nicola}, \citenamefont {Ruiz De~Elvira},\ and\ \citenamefont
  {Vioque-Rodr\'\i{}guez}}]{GomezNicola:2019myi}%
  \BibitemOpen
  \bibfield  {author} {\bibinfo {author} {\bibfnamefont {A.}~\bibnamefont
  {G\'omez~Nicola}}, \bibinfo {author} {\bibfnamefont {J.}~\bibnamefont {Ruiz
  De~Elvira}}, \ and\ \bibinfo {author} {\bibfnamefont {A.}~\bibnamefont
  {Vioque-Rodr\'\i{}guez}},\ }\href {\doibase 10.1007/JHEP11(2019)086}
  {\bibfield  {journal} {\bibinfo  {journal} {JHEP}\ }\textbf {\bibinfo
  {volume} {11}},\ \bibinfo {pages} {086} (\bibinfo {year} {2019})},\ \Eprint
  {http://arxiv.org/abs/1907.11734} {arXiv:1907.11734 [hep-ph]} \BibitemShut
  {NoStop}%
\bibitem [{\citenamefont {Zyla}\ \emph {et~al.}(2020)\citenamefont {Zyla} \emph
  {et~al.}}]{Zyla:2020zbs}%
  \BibitemOpen
  \bibfield  {author} {\bibinfo {author} {\bibfnamefont {P.~A.}\ \bibnamefont
  {Zyla}} \emph {et~al.} (\bibinfo {collaboration} {Particle Data Group}),\
  }\href {\doibase 10.1093/ptep/ptaa104} {\bibfield  {journal} {\bibinfo
  {journal} {PTEP}\ }\textbf {\bibinfo {volume} {2020}},\ \bibinfo {pages}
  {083C01} (\bibinfo {year} {2020})}\BibitemShut {NoStop}%
\bibitem [{\citenamefont {Gasser}\ and\ \citenamefont
  {Leutwyler}(1984)}]{Gasser:1983yg}%
  \BibitemOpen
  \bibfield  {author} {\bibinfo {author} {\bibfnamefont {J.}~\bibnamefont
  {Gasser}}\ and\ \bibinfo {author} {\bibfnamefont {H.}~\bibnamefont
  {Leutwyler}},\ }\href {\doibase 10.1016/0003-4916(84)90242-2} {\bibfield
  {journal} {\bibinfo  {journal} {Annals Phys.}\ }\textbf {\bibinfo {volume}
  {158}},\ \bibinfo {pages} {142} (\bibinfo {year} {1984})}\BibitemShut
  {NoStop}%
\end{thebibliography}%
\end{document}